\documentclass[showpacs,preprintnumbers,amsmath,amssymb,twocolumn,superscriptaddress]{revtex4-1}
\usepackage{epsfig} 
\usepackage{graphicx, latexsym, amssymb, amsmath, color, multirow, mathrsfs, booktabs}
\usepackage{dcolumn}
\usepackage{bm}
\usepackage{soul}

\newcommand{\frc}[2]{\mbox{$\frac{#1}{#2}$}}

\begin{document}

\title{Densities and energies of nuclei in dilute matter} 
\author{P.~Papakonstantinou}
\affiliation{Institut de Physique Nucl\'eaire, IN2P3-CNRS, Universit\'e Paris-Sud, 91406 Orsay, France.}
\author{J.~Margueron}
\affiliation{Institut de Physique Nucl\'eaire, IN2P3-CNRS, Universit\'e Paris-Sud, 91406 Orsay, France.}
\affiliation{Institut de Physique Nucl\'eaire, IN2P3-CNRS, Universit\'e Lyon 1, 69622 Villeurbanne, France.}
\author{F.~Gulminelli}
\affiliation{CNRS, ENSICAEN, UMR6534, LPC ,F-14050 Caen cedex, France}
\author{Ad.R.~Raduta}
\affiliation{NIPNE, Bucharest-Magurele, POB-MG6, Romania}

\date{\today}

\begin{abstract}
We explore the ground-state properties of nuclear clusters embedded in a gas of nucleons with the help of Skyrme-Hartree-Fock microscopic calculations. 
Two alternative representations of clusters are introduced, namely coordinate-space and energy-space clusters. 
 We parameterize their density profiles in spherical symmetry in terms of basic properties of the energy density functionals used and propose an analytical, Woods-Saxon density profile whose parameters depend, not only on the composition of the cluster,  but also of the nucleon gas. 
We study the clusters' energies with the help of the local-density approximation, validated through our microscopic results.  
We find that the volume energies of {\em coordinate-space} clusters are determined by the saturation properties of matter, while the surface energies are strongly affected by the presence of the gas. 
We conclude that both the density profiles and the cluster energies are strongly affected by the gas and discuss implications for the nuclear EoS and related perspectives. 
Our study provides a simple, but microscopically motivated modeling of the energetics of clusterized matter at subsaturation densities, for direct use in consequential applications of astrophysical interest. 
\end{abstract}

 
 \maketitle


\section{Introduction}

The understanding of the
structure and evolution of neutron stars, and the cataclysmic events generating them, 
is strongly related to our knowledge of
the microphysics of dense matter.
Additionally, the
implementation 
of the equation of state (EoS) 
in compact-star 
modeling
presents a double challenge: 
First, knowledge of the structure and dynamics of hadronic matter over different domains of temperature, density and composition is needed - including important phase transitions and interactions with leptons and photons; 
second, this knowledge must be weaved together and modeled 
in such a way, that numerical studies and simulations are feasible. 
Terrestrial experiments on nuclei, including systems under extreme conditions, are performed and analysed worldwide in order to meet the former challenge. It remains a non-trivial task to model and tabulate the acquired information into usable form for astrophysical applications. 

In terms of relevant degrees of freedom, matter at low temperature and at sub-saturation density is not highly exotic: it will primarily consist of nucleons, leptons, and photons and therefore belong in the domain of low-energy nuclear physics. 
Depending on the situation, baryonic matter may be almost isospin-symmetric or extremely neutron rich - as in the inner crust of a neutron star. 
A liquid-gas phase transition is expected to occur at sub-saturation densities, from 
non-uniform
matter to unbound nucleons. 
Of particular interest in the present work is the intermediate domain of clusterized matter, where nuclei - as it were, nuclear droplets - coexist with free nucleons, as well as leptons and photons, in thermal and chemical equilibrium. 
Again depending on the conditions, different nuclei may be found in this mixture: from very light ones, like deuterons and alpha particles, to iron-group nuclei, and much heavier or exotic species. 

A common approach to the composition of clusterized stellar matter, followed in the most widely used 
EoS's, namely those by J.M.~Lattimer and F.D.~Swesty~\cite{Lattimer1991} or H.~Shen et al.~\cite{Shen1998}, is the so-called single-nucleus approximation (SNA): 
Besides free nucleons, only one kind of light cluster (alpha particles) and one kind of heavy cluster are assumed to exist. 
The idea is to account in an average way for the properties of the 
statistical
cluster distribution. 
The SNA may not affect very strongly thermodynamical properties of matter at certain temperature and 
density domains of interest~\cite{Burrows1984}, 
but it may have consequences for dynamical processes dependent on reaction rates of specific nuclei~\cite{Langanke2003}
and for the gas-liquid phase transition.  
Therefore, more modern approaches rely on an extended nuclear statistical equilibrium (NSE) concept, where the distribution of clusters over, in principle, all mass numbers is taken into account and obtained self-consistently in conditions of statistical equilibrium~\cite{Phillips1994,Botvina2010,Souza2009}.

Originally, the NSE was introduced to describe the reaction network taking place at the end of the evolution
of massive stars in red supergiants.
Being very diluted, nuclei interact weakly and are almost not modified by the surrounding medium.
These conditions naturally lead to the Saha equations.
The NSE in the dense and hot matter in the core of supernovae was first applied in the EoS of Hillebrandt and
Wolff~\cite{Hillebrandt1984}.
In recent NSE implementations~\cite{Hempel2010,Raduta2010,Hempel2011,Gulminelli2012}   
the interactions  
between a cluster and the surrounding gas is treated in the so-called excluded-volume approach. 
The clusters and the gas of light particles do not overlap in space and the clusters 
binding energy is kept as in the free limit.
It is known, however, from quantal approaches, that the cluster properties are 
modified
by the coexistence with a gas~\cite{Typel2010a}. 
Such approaches are presently limited to 
a system
of light nuclei. It is clear that  in-medium modifications of larger clusters  would also change the composition of matter and the related EoS. It is therefore important to take such effects into account. 

This work provides a microscopically based modelling
of in-medium properties of nuclei, specifically their ground-state densities and energies. We focus on medium-mass and heavy clusters. 
Clusters are first studied microscopically through Hartree-Fock calculations in Wigner-Seitz cells. 
Clustering emerges from the quantal calculation since a dense cluster can be distinguished from a dilute
phase.
Two different types of clusters can however be defined:
a representation in coordinate space, pertinent in excluded-volume approaches; and another representation
in energy space, conceptually similar to existing quantal approaches for light clusters. 
Next, we study the energies of clusters embeded in a gas, 
with the help of the Local-Density Approximation (LDA), which we validate using our microscopic results. 
Thanks to our analytical model for density and the LDA, configurations away from those accessible within Hartree-Fock or other variational calculations can be studied and the in-medium modified self-energies can be directly used in the EoS modelling at finite temperature. 

For the microscopic calculations we employ the SLy4 Skyrme functional, but our 
in-medium densities and energies explicitly depend on the isoscalar and isovector 
properties of the underlying interaction, and can thus readily be computed for different functionals. 

The outline of our paper is as follows. 
In Sec.~\ref{sec:theor}, basic information is provided on our theoretical methods, namely the Skyrme-Hartree-Fock model, the LDA and the Woods-Saxon functions. 
In Sec.~\ref{paperpana1:sec:analysis}, we analyse our microscopic results and define two types of clusters. 
In Sec.~\ref{sec:anmodden}, we present and validate our analytical model for the density profiles of clusters in a gas. 
In Sec.~\ref{sec:energy} we study their energies within the LDA. 
We conclude in Sec.~\ref{sec:concl}.

\section{Theoretical methods and considerations\label{sec:theor}} 

\subsection{Microscopic calculations 
\label{sec:microcalc}} 

Our modelization of clusters embedded in a nucleon gas will initially be informed by microscopic Hartree-Fock calculations using Skyrme functionals. 
The Hartree-Fock equations are solved self-consistently in spherical symmetry in a mesoscopic volume, 
which can be associated with a Wigner-Seitz cell. 
Its radius $R_{cell}$ must be much larger than that of a typical nucleus. 
Here we will use  $R_{cell}=35$~fm. 
Implicit here is the assumption, that the clusters are sufficiently far from each other to not interact. 
The purpose of the mesoscopic volume in its present use is to allow for the formation of a uniform gas around the cluster. 
Its radius is 
not chosen via a condition of charge neutrality. 
It is simply an auxiliary quantity which will be eliminated from the final results. 

The equations are solved with mixed boundary conditions: 
Dirichlet boundary conditions are imposed on odd single-particle states and von Neumann conditions  on even states. 

In principle, Hartree-Fock-Bogolyubov calculations could be performed, in order to include pairing~\cite{Dobaczewski:1983zc}. 
Superfluidity is an essential property of dilute star matter~\cite{Baldo2007,Grill2011,Pearson2011}, 
but we expect the in-medium modifications due to pairing to be small in comparison with the in-medium modifications due to the mean-field. As we will show, these latter can be recasted into in-medium modified bulk and 
surface terms of the cluster energy functional. To correctly constrain such terms from the microscopic calculations and disentangle them from possible pairing terms, we therefore prefer to neglect pairing and stick to a Hartree-Fock modeling.

The energy density for a traditional Skyrme functional is given by~\cite{Ducoin2006}
\begin{eqnarray} 
  \mathcal{H}(r) &=& 
\frac{\hbar^2}{2m_p}\tau_p + \frac{\hbar^2}{2m_n}\tau_n  
\nonumber \\   && 
+ C_0 \rho^2
+ C_3 \rho^{\alpha + 2 } 
+ C_{\mathrm{eff}} \rho\tau 
\nonumber \\   && 
+ D_0 \rho_3^2 
+ D_3 \rho^{\alpha} \rho_3^2 
+ D_{\mathrm{eff}} \rho_3\tau_3 
\nonumber \\   && 
+ C_{12} \rho\nabla^2\rho 
+ D_{12} \rho_3\nabla^2\rho_3  
\nonumber \\   && 
-\frac{W_0}{4}[3\rho\vec{\nabla}\cdot\vec{J} + \rho_3\vec{\nabla}\cdot\vec{J}_3]  
,  
\label{eq:skyrmeHF} 
\end{eqnarray}  
where 
\begin{eqnarray*}
&& C_0 = \frc{3}{8}t_0 
\quad 
; 
\quad 
C_3 = \frc{1}{16}t_3 
\\ 
&& C_{\mathrm{eff}} = \frc{1}{16}[3t_1 
+ t_2 (4x_2 + 5)] 
\\ 
&& D_0 = -\frc{1}{8}t_0(2x_0 + 1) 
\quad 
; 
\quad 
D_3 = -\frc{1}{48}t_3(
2 x_3+1) 
\\ 
&& D_{\mathrm{eff}} = \frc{1}{16} [  -t_1(2x_1  
+  1) + t_2 (2x_2+1)  ]  
\\ 
&&C_{12} = \frc{1}{64} [-9t_1 + t_2 (5+4x_2) ] 
\\ 
&&D_{12} = \frc{1}{32} [3t_1(\frc{1}{2}+x_1) + t_2 (\frc{1}{2}+x_2)] 
\end{eqnarray*}  
and $t_i$, $x_i$ are the usual Skyrme parameters. 
In the above,  
\begin{equation} 
\rho = \rho_p + \rho_n \quad ; \quad \rho_3 = \rho_n - \rho_p 
\label{eq:rhoHF} 
\end{equation} 
are the isoscalar and isovector densities, while 
\begin{equation} 
\tau = \tau_p + \tau_n \quad ; \quad \tau_3 = \tau_n - \tau_p 
\label{eq:tauHF} 
\end{equation} 
give the isoscalar and isovector expectation-value densities of the momentum-squared operator  
and  
\begin{equation} 
\vec{J} = \vec{J}_p + \vec{J}_n \quad ; \quad \vec{J}_3 = \vec{J}_n - \vec{J}_p 
\label{eq:JHF} 
\end{equation} 
are the current densities. 
The densities are functions of the radial variable $r$ and the subscript $p$ or $n$ specifies the densities of protons or neutrons, respectively.

The SLy4 Skyrme functional~\cite{Chabanat1998a,Chabanat1998b} has been chosen at present for the microscopic calculations. 
This functional is widely used to calculate ground and excited states of nuclei 
and it is particularly interesting in studies of dense matter, 
because is has been adjusted to a realistic EoS of symmetric and neutron matter, 
obtained variationally using the UV14 and UVII potentials~\cite{Wiringa1988,Wiringa1993}.
We stress, however, that our modeling will be general in scope, so that it can be easily applicable to different functionals and interactions. 

In practice we will study isotopic chains ranging from $Z=20$ (Ca) to $Z=126$ (Pb). 
For each $Z$ we vary the neutron number from 
somewhat below
 $Z$ to approximately 3000. 
We thus obtain a variety of solutions, from stable nuclei to very neutron-rich clusters embedded in a neutron gas 
of density up to, approximately, 0.02~fm$^{-3}$. 
We are therefore able to study microscopically neutron-rich clusterized matter, before the onset of deformation and
pasta phases. 
The energies and density profiles obtained from the microscopic calculations will be used to validate our analytical models.

The Coulomb interaction among protons in stellar matter is screened by the neutralizing electron background.
The associated modification to the cluster Coulomb energies is an important in-medium effect which is already accounted for in all EoS-NSE based models~\cite{Hempel2010,Raduta2010}. A possible extra effect of Coulomb screening on the nuclear part of the energy functional through a modification of the density profile is expected to be small  in the very neutron rich nuclei that constitute our microscopic sample~\cite{Bonche1985}. 


\subsection{Energy in the local-density approximation} 

Within the Local-Density Approximation (LDA), the local energy density is determined by the proton and neutron densities and their gradients, while current densities are ignored. 
The LDA is closely related to the Thomas-Fermi approach whereby the ground-state properties can be determined by minimizing the energy with respect to variations of the density. 
For our purposes, using the LDA will 
transpose
the problem of determining the cluster energy in-medium correction towards the determination of the nuclear density 
profile in the Wigner-Seitz cell.

Within the LDA, the kinetic energy density is also written in terms of the local density, being replaced by its value in nuclear matter of the corresponding density. 
In spherical symmetry we thus write, 
\begin{equation} 
\tau_{p,n}(r) = \frc{3}{5} \rho_{p,n} k_{\mathrm{F} \, p,n}^2(r) 
, 
\end{equation} 
where $k_{\mathrm{F} \, p,n} = (3\pi^2\rho_{p,n})^{1/3}$ is the Fermi momentum of protons and neutrons in infinite matter of density $\rho_{p,n}$.  
With these considerations, the energy density in the LDA~(\ref{eq:skyrmeHF}) is given by  
\begin{eqnarray} 
  \mathcal{H}_{LDA}(r) &=& 
\frac{\hbar^2}{2m_p}\tau_p    
+ \frac{\hbar^2}{2m_n}\tau_n  
\nonumber \\   && 
 +  C_0 \rho^2 
+ C_3 \rho^{\alpha + 2 } 
+ C_{\mathrm{eff}} \rho\tau 
\nonumber \\   && 
+ D_0 \rho_3^2 
+ D_3 \rho^{\alpha} \rho_3^2 
+ D_{\mathrm{eff}} \rho_3\tau_3 
\nonumber \\   && 
+ C_{12} \rho\nabla^2\rho 
+ D_{12} \rho_3\nabla^2\rho_3  
,  
\label{eq:skyrmeLDA} 
\end{eqnarray}  
where 
\begin{equation} 
\tau_{p,n} = \frc{3}{5} (3\pi^2)^{2/3} \rho_{p,n}^{5/3} 
\end{equation} 
and $\tau , \tau_3$ are given by Eq.~(\ref{eq:tauHF}). 
The total energy in a spherical cell of radius $R_{cell}$ will be given simply by 
\begin{equation} 
E_{LDA} = 4\pi \int_0^{R_{cell}} \mathcal{H}_{LDA}(r) r^2 \mathrm{d}r  
. 
\end{equation} 
All that is needed for computing the energy of the system is then its density profile - both the isoscalar and isovector components. 

\subsection{Nuclear density profiles \label{sec:nucdenprof}} 

Density profiles of medium mass and heavy nuclei are  known to be  well described 
by Woods-Saxon profiles, 
\begin{equation} 
\rho^{WS} (r) = \frac{\rho_0}{1 + \exp \left[ (r-R^{WS})/a^{}\right] }
\end{equation} 
where $R^{WS}$ is the radius and $a$ is the diffuseness parameter of the Woods-Saxon profile. 
Other analytical forms have been considered in the literature~\cite{Friedrich1982pp,Andrae2000pp}
and results marginally depend on the chosen form.
We note that Hartree-Fock calculations yield density profiles with small ripples in the bulk, 
but those are to a large extent artefacts of the mean-field approach 
and are expected to be washed out by correlations, with the obvious exception of exotic bubble-shaped nuclei~\cite{Khan200837}. 

A straighforward generalization of the above expressions for nuclei in the presence of 
a homogeneous gas with density $\rho_{gas}$ 
in spherical symmetry is given by: 
\begin{eqnarray}
\rho^{WS}_{cell} (r) &=&  \frac{\rho_0}{1 + \exp \left[ (r-R^{WS})/a^{}\right] } 
  \nonumber \\ 
 & & + \frac{\rho_{gas}}{1 + \exp \left[- (r-R^{WS})/a^{}\right] }
 \nonumber \\ 
 &=&  \frac{\rho_0-\rho_{gas}}{1 + \exp \left[ (r-R^{WS})/a^{}\right] } + \rho_{gas}  
\label{eq:rhototWS} 
\end{eqnarray} 
It is straightforward to show that 
\begin{equation}
 a^{}=-\frac{\rho_0 - \rho_{gas}}{4\rho'(R^{WS})}
\label{eq:adiffWS} 
\end{equation}
and 
\begin{equation} 
\rho^{WS}_{cell}(R^{WS}) = \frac{\rho_0+\rho_{gas}}{2} 
. 
\end{equation} 
Analogous expressions may be introduced separately for the proton and neutron densities. 

The density profiles of ordinary nuclei existing on earth or produced in nuclear facilities 
have been analyzed before in semi-classical Thomas-Fermi models in terms of Woods-Saxon functions (see for instance 
Ref.~\cite{Brack1985}), 
corresponding in our case to setting 
the external density $\rho_{gas}$ to zero.
Application of the Extended Thomas-Fermi model for dilute nuclei in neutron stars has also been considered
on the basis of Woods-Saxon functions in Ref.~\cite{Onsi2008}.
In the case where only one kind of particles is dripping out of the cluster, the neutrons for instance, the external 
proton density $\rho_{gas,p}$ is zero. 
In this work we mostly analyze neutron rich nuclei where $\rho_{gas,p}=0$.
However, our approach can also be applied in cases where there are protons in the gas.

\section{Analysis of the density profile
\label{paperpana1:sec:analysis}}

In this Section we analyze the density profiles obtained microscopically and we
define two kinds of representations for the in-medium clusters.
Simple relations between these two representations will also be derived.
At first we fit the microscopic Hartree-Fock profiles with Woods-Saxon functions, 
in order to determine useful parameters, e.g., the bulk density and asymmetry. 

\begin{figure*} [t]
\includegraphics[width=0.7\textwidth,angle=-90]{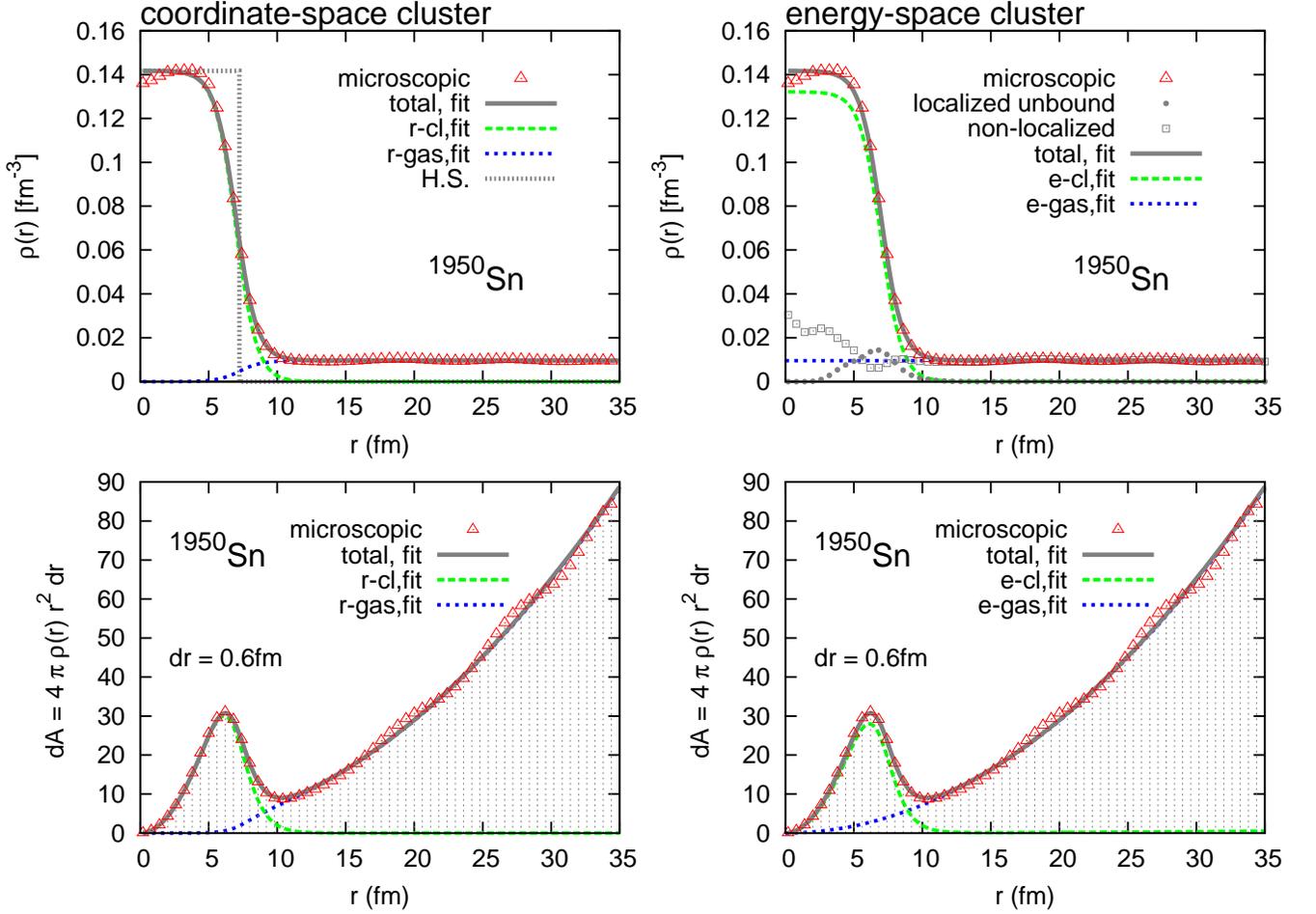} 
\caption{{\em Top:} A typical microscopic density profile (triangles), corresponding to 50 protons and 1900 neutrons. 
The solid grey line corresponds to the Woods-Saxon fit of the total density, the green dashed line to the cluster density (left: r-cluster; right: e-cluster) and the blue short-dashed line to the gas. 
The density of unbound nucleons, localized or non-localized, is also shown. 
{\em Bottom:} Number of particles in volume elements as indicated, for the density profiles shown on top. 
The vertical lines indicate radial elements $dr = 0.6$~fm. 
\label{fig:SnN1900rho}} 
\end{figure*} 

The microscopic density profile of nucleonic matter $\rho_{cell}(r)$ is represented on the top part of
Fig.~\ref{fig:SnN1900rho} in the specific case of $^{1950}$Sn (triangles). 
On the bottom part of the figure we show also the profile of the number of particles $4\pi \rho_{cell}(r) r^2 d r$ 
from which is obtained the total number of particles in the cell,
\begin{equation}
A_{cell} = 4\pi \int_0^{R_{cell}} r^2 \rho_{cell}(r) \, dr.
\end{equation}
It is clearly visible from the top part of Fig.~\ref{fig:SnN1900rho} (left and right) that there is a dense nuclear cluster in the center of  the cell and a dilute neutron gas in its outer part. 
However, the 
central
 cluster contains only a fraction of the number of particles as it is shown
on the bottom part of Fig.~\ref{fig:SnN1900rho} and in this case, most of the particles are in the 
non-central region.

There is an ambiguity on how to define dilute clusters, related to the overlap of the cluster 
with the gas: does the external gas penetrate inside the cluster, 
or is it excluded from the cluster? 
These two representations of the same system are depicted in Fig.~\ref{fig:SnN1900rho} where
the two panels on the left show how to build up the microscopic cell density $\rho_{cell}(r)$ from 
two excluded sub-densities, while the two panels on the right show how the penetration of the
external gas inside the cluster can be taken into account.
We name these two representations the coordinate-space clustering 
(left of Fig.~\ref{fig:SnN1900rho}) and the energy-space clustering (right of Fig.~\ref{fig:SnN1900rho}), 
respectively.
The coordinate-space clustering is the representation of clusters which naturally emerges
if the density is taken as the relevant degree of freedom, as in the density functional theory. 
In this picture, clusters are recognized as density fluctuations on top of a homogeneous medium,  
and occupy a volume in space. 

In the following, we aim at providing a global model for the density profile, but first, we shall discuss
the two types of clustering. 

\subsection{Coordinate-space clustering}

Clusters defined in coordinate space occupy a volume in space and are surrounded by a gas of light particles. 
In the following, these clusters will be referred to as the \textsl{r-clusters} and the gas as the \textsl{r-gas}.
The frontier between the \textsl{r-cluster} and the \textsl{r-gas} is located in space at the place where
the nucleonic density changes drastically.
Hereafter, it will be referred to as the surface of the \textsl{r-cluster}.
Since the \textsl{r-cluster} has a surface diffuseness, the frontier cannot be sharp.
At the surface of the \textsl{r-cluster}, there is therefore a narrow region in space where the \textsl{r-cluster}
and the \textsl{r-gas} overlap, with a typical size of 1-2~fm, as illustrated on the left part of 
 Fig.~\ref{fig:SnN1900rho}.

The coordinate-space clusters have been mostly used to set up semi-classical EoS's such as
the Lattimer-Swesty EoS~\cite{Lattimer1991} or other modelings of the crust of neutron stars. 
We note that in the semi-classical EoS~\cite{Lattimer1991}, the separation between the \textsl{r-cluster} and the
\textsl{r-gas} is sharp in coordinate space, 
but it is not so in the microscopic calculations.
The idea of clusters as spatial structures in Wigner-Seitz cells was already put forth in the earliest studies of clusterized matter~\cite{Negele1973}. 
We propose in the following a method to relate this semi-classical modeling with the microscopic calculations
where we use Woods-Saxon functions to fit the microscopic density in the cell.

In the \textsl{r-clustering} representation, the density profile of the nucleons 
represented 
in Fig.~\ref{fig:SnN1900rho} is decomposed in terms of an inside Woods-Saxon function, 
\begin{equation}
\rho^{WS}_{r-cl}(r)= \frac{\rho_{0}}{1+\exp [(r-R^{WS})/a^{}]} ,
\label{paperpana1:eq:ws1}
\end{equation}
surrounded by an outside gas density defined as
\begin{equation}
\rho^{WS}_{r-gas}(r)= \frac{ \rho_{gas}}{1+\exp [-(r-R^{WS})/a^{}]},
\label{paperpana1:eq:ws2}
\end{equation}
so that the sum of the two profiles gives the total profile $\rho_{cell}^{WS}(r)$, introduced in Eq.~(\ref{eq:rhototWS}). 
The parameters of the 
Woods-Saxon density profile are: the \textsl{r-cluster} bulk density $\rho_0$, the radius $R^{WS}$, the 
surface diffuseness $a^{}$, and the asymptotic gas density $\rho_{gas}^{WS}$.
The same quantities can be introduced separately for neutrons and protons, namely
\begin{equation}
\rho_{r-cl,q}^{WS}(r) = \frac{\rho_{0,q}}{1+\exp [(r-R_q^{WS})/a_q^{WS}]} ,
\label{paperpana1:eq:ws1q}
\end{equation}
and
\begin{equation}
\rho_{r-gas,q}^{WS}(r) = \frac{ \rho_{gas,q}}{1+\exp [-(r-R_q^{WS})/a_q]}.
\label{paperpana1:eq:ws2q}
\end{equation}
In Eqs.~(\ref{paperpana1:eq:ws1q})-(\ref{paperpana1:eq:ws2q}) the index $q$ stands for neutrons ($n$) or protons ($p$.) 
The Woods-Saxon parameters entering Eqs.~(\ref{paperpana1:eq:ws1})-(\ref{paperpana1:eq:ws2q}) can be obtained with the help of fits on microscopic nucleon, neutron and proton density profiles, 
as we describe in the Appendix. 
An analytical model for the parameters will be introduced in Sec.~\ref{sec:anmodden}. 

Since the sum of two Woods-Saxon functions with different radii and diffuseness parameters is no longer a Woods-Saxon function, only two of the three density profiles defined above (for protons, neutrons and nucleons) can be employed simultaneously. In this work we consider the nucleon and proton densities as the independent functions. 
The nucleon and proton densities in the cell are defined as,
\begin{equation}
\rho_{cell}^{WS}(r) = \rho^{WS}_{r-cl}(r)+\rho^{WS}_{r-gas}(r),
\label{paperpana1:eq:rhocell}
\end{equation} 
and
\begin{equation}
\rho_{cell,p}^{WS}(r) = \rho_{r-cl,p}^{WS}(r)+\rho_{r-gas,p}^{WS}(r),
\label{paperpana1:eq:rhospacep}
, 
\end{equation} 
respectively, 
while for the neutrons we choose 
\begin{equation} 
\rho_{cell,n}^{WS} (r) = \rho_n^{WS (1)} (r) = \rho^{WS}(r) - \rho_p^{WS}(r) 
\label{paperpana1:eq:rhospacen1}
\end{equation} 
to be compared with 
\begin{equation} 
\rho_{cell,n}^{WS (2)} = \rho_{r-cl,n}^{WS}(r)+\rho_{r-gas,n}^{WS}(r)
\label{paperpana1:eq:rhospacen2}
\end{equation}
obtained from a separate fit on the neutron density.

On the left of Fig.~\ref{fig:SnN1900rho} is shown the result of the fit for the nucleon 
density~(\ref{paperpana1:eq:rhocell}) in solid line,
as well as the contributions of the cluster density $\rho^{WS}_{r-cl}(r)$ and gas density 
$\rho^{WS}_{r-gas}(r)$ represented respectively by the dashed and dotted lines. 

In the following 
we drop the label ``WS" from our notations. 

The number of nucleons in the \textsl{r-cluster},  $A_{r-cl}$, and the numbers of nucleons in the gas, $A_{r-gas}$, are defined as
\begin{equation}
A_{r-cl} = \int d^3r  \rho_{r-cl}(r) ,
\quad , \quad 
A_{r-gas} = \int d^3r  \rho_{r-gas}(r).
\end{equation}
We have similarly for the number of protons in the cluster, $Z_{r-cl}$, and in the gas, $Z_{r-gas}$, 
\begin{equation}
Z_{r-cl} = \int d^3r  \rho_{r-cl,p}(r) ,
\quad , \quad 
Z_{r-gas} = \int d^3r \rho_{r-gas,p}(r),
\end{equation}
the latter being zero in the case of Fig.~\ref{fig:SnN1900rho}. 
Finally, the number of neutrons in the cluster, $N_{r-cl}$, and in the gas, $N_{r-gas}$, are 
\begin{equation} 
\begin{array}{c}  
N_{r-cl} = N_{r-cl}^{(1)} = A_{r-cl} - Z_{r-cl} \\ 
N_{r-gas} = N_{r-gas}^{(1)} = A_{r-gas}-Z_{r-gas} . 
\end{array} 
\label{paperpana1:eq:nrcl1} 
\end{equation} 
Alternatively, 
\begin{equation} 
\begin{array}{c}  
N_{r-cl}^{(2)} = \int d^3r  \rho_{r-cl,n}(r) ,
\\ 
N_{r-gas}^{(2)} = \int d^3r \rho_{r-gas,n}(r),
\end{array} 
\label{paperpana1:eq:nrcl2}
\end{equation} 
from a separate fit on the neutron density. 

We have checked that our numerical results, for which there is no proton gas, are practically the same (differences by less than 1\%) if we treat the neutron distribution as independent or as the difference of the total and the proton ones. 
We chose to work with the difference (i.e., with only two independent density distributions) 
to ensure exact conservation of the total number of particles. This way we also avoid ambiguities in the definition of the cluster volume. 
 
\subsection{Energy-space clustering}

The energy-space clustering is an alternative  representation of the density profile where
clusters are defined as a collection of bound particles, independent of their localization. 
As a consequence, the external gas is allowed to penetrate inside the cluster.
At variance with r-clustering, this representation allows extending the concept of clustering to supercritical densities~\cite{Sator:2002br}. 

This representation appears natural in quantum mechanics,  where wave-functions with various quantum numbers can overlap.
Since the separation between the cluster and the gas is described in terms of quantum single-particle wave-functions
mainly characterized by their energies, as we discuss below,
we refer to these two ensembles of particles as the \textsl{e-cluster} and the \textsl{e-gas}.

Before we discuss the microscopic origin of \textsl{e-clustering}, we begin with an analytical parameterization of the \textsl{e-cluster} and \textsl{e-gas} density profiles. The density of the \textsl{e-cluster} is defined as
\begin{equation}
\rho^{WS}_{e-cl}(r) = \frac{\rho_{0}-\rho_{gas}}{1+\exp [(r-R^{WS})/a^{}]} ,
\label{paperpana1:eq:wscluster}
\end{equation}
and the \textsl{e-gas} density, $\rho^{WS}_{e-gas}(r)$, is set to be constant in the coordinate variable $r$,
\begin{equation}
\rho^{WS}_{e-gas}(r) = \rho_{gas} .
\label{paperpana1:eq:wsgas}
\end{equation}
The various parameters are the same as in the case of \textsl{r-clustering}.
It is readily verified that the total density in the energy representation, 
\begin{equation} 
\rho_{cell}^{WS}(r) = \rho^{WS}_{e-cl}(r) + \rho^{WS}_{e-gas}(r) ,
\label{paperpana1:eq:rhocelle} 
\end{equation} 
is the same as 
in the case of \textsl{r-clustering} and yields Eq.~(\ref{eq:rhototWS}). 
Therefore, the fit does not depend on the representation. 
Comparing Eqs.(\ref{paperpana1:eq:ws1}) and (\ref{paperpana1:eq:wscluster}), we obtain
\begin{equation} 
\rho^{WS}_{e-cl}(r) = \left( 1 - \frac{\rho_{gas}}{\rho_{0}}\right) \rho^{WS}_{r-cl}(r) 
\label{paperpana1:eq:re-prop}
, 
\end{equation} 
implying 
\begin{equation}
A_{e-cl} = \left(1-\frac{\rho_{gas}}{ \rho_{0}} \right) A_{r-cl} 
\label{paperpana1:eq:relation},
\end{equation}
where $A_{e-cl}$ is defined hereafter.

Completely analogous densities $\rho_{e-cl,q}^{WS}(r)$ and related quantities can be defined for the neutron and proton densities separately. 
The neutron density profile and the number of neutrons in the \textsl{e-cluster} can be obtained in two ways, as in the case of the \textsl{r-cluster}, 
cf. Eqs.~(\ref{paperpana1:eq:nrcl1}),~(\ref{paperpana1:eq:nrcl2}). 
Here we adopt 
\begin{equation} 
\begin{array}{c} 
 N_{e-cl} = N_{e-cl}^{(1)} = A_{e-cl} - Z_{e-cl}  
\\ 
N_{e-gas} = N_{e-gas}^{(1)} = A_{e-gas}-Z_{e-gas} . 
\end{array} 
\label{paperpana1:eq:necl1}
\end{equation} 
in analogy to 
Eq.~(\ref{paperpana1:eq:nrcl1}). 

The density profiles~(\ref{paperpana1:eq:wscluster}) and (\ref{paperpana1:eq:wsgas}) are represented in
the right panels of Fig.~\ref{fig:SnN1900rho} by the dashed and dotted line respectively. The
nucleon density in the cell obtained from the fit is represented by the solid line and is identical to the one on the left panel. 
Equations (\ref{paperpana1:eq:re-prop}), (\ref{paperpana1:eq:relation}), (\ref{paperpana1:eq:necl1}) show that there is an exact and analytical mapping between the two representations, and we will therefore use both in the following.

To get an insight on the physical meaning of e-clustering, 
it is possible to extract the density associated to the 
\textsl{bound} states, the \textsl{resonant} states and the \textsl{continuum} states from the microscopic quantum calculation.
We have classified the wave functions into three groups depending on their single-particle energies and
space-extension inside the cell.
First we define the quantity $u_{mf}^{ext}$ as the external mean field, at a large distance from the cluster.
A \textsl{bound} wave function is considered to have, by definition, a single-particle energy $\epsilon_i < u_{mf}^{ext}$, otherwise
the wave-function belongs either to the \textsl{resonant} states or to the \textsl{continuum} states.
Since in the microscopic calculation the states with $\epsilon_i > u_{mf}^{ext}$ are expanded in a discreet-box basis, 
it is difficult to unambiguously distinguish the \textsl{resonant} states from the \textsl{continuum} states.
In our case they are differentiated according to their coordinate space extension: unbound wave-functions populating substantially the cell outside the cluster, 
with an amplitude comparable (within $30$\%) to normalized plane waves,  
are \textsl{non-nocalized} and are set to be in the \textsl{continuum}, 
otherwise 
they 
are considered \textsl{resonant} states.
The accuracy of this geometric method is limited by the wave-functions close to the 
centrifugal barrier which remain difficult to identify.
There is an ambiguity of few nucleons in this method which is enough for our discussion.

On the right side of Fig.~\ref{fig:SnN1900rho} we show the density profile of the continuum, or \textsl{non-localized} unbound states (squares) and 
that of the \textsl{resonant} or \textsl{localized} unbound states (dots).
It is interesting to notice that the density due to continuum states penetrates well inside the cluster, which confirms
that the gas is not excluded from the central part of the cell.
Additionally, one can notice that the continuum-state density slightly increases inside the cluster compared
to the asymptotic density.
This property shall be related to the attractive nature of the proton-neutron interaction.
The simpler approximation for the density of the gas inside the cluster is simply to set it constant and defined
by the asymptotic density, as anticipated with  
Eq.~(\ref{paperpana1:eq:wsgas}).  
The error introduced in the population of the cluster is small. 

\begin{figure*} [t]
\includegraphics[width=0.6\textwidth,angle=-90]{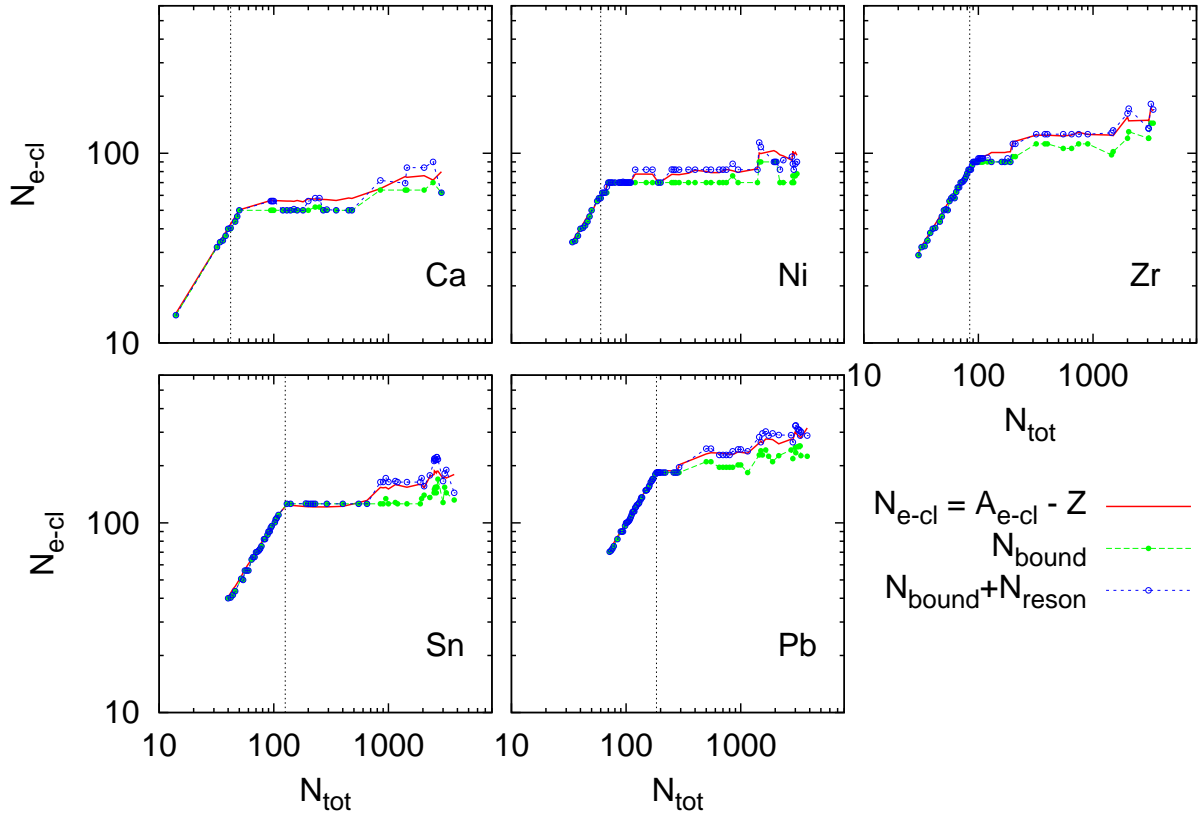} 
\caption{For given number of protons (isotopic chain), and as a function of total amount of neutrons in the cell, 
shown are: the number of neutrons in the energy-space cluster, computed as the integral of the \textsl{e-cluster} density profile and assuming a uniform gas, $N_{e-cl}$; 
the number of neutrons in bound states, $N_{bound}$; 
and the number of neutrons in bound or in ``resonant", i.e., localized unbound states, $N_{bound}+N_{reson}$. 
\label{paperpana1:fig:NclNtot}} 
\end{figure*}

The density of the \textsl{resonant}, or \textsl{localized}, unbound states is also shown in Fig.~\ref{fig:SnN1900rho}. 
There is a clear overlap in coordinate-space between the density of the resonant states and the cluster.
If we want to define an e-cluster in terms of its single particle states, an ambiguity exists concerning resonant states. 
These could be attributed to the gas, 
since they are unbound,
or to the cluster, since in the e-clustering concept the gas is constituted by an homogeneous background, i.e., by single particle plane waves, 
while resonant states are clearly affected by the cluster mean-field.  
In order to reach a  decision, we analyze further our microscopic calculations on a large number of systems in Fig.~\ref{paperpana1:fig:NclNtot}. 
The number of neutrons in the energy-space cluster $N_{e-cl}$ is first estimated from the fit of the cell density 
and proton density via Eq.~(\ref{paperpana1:eq:necl1}). 
$N_{e-cl}$ is represented by the red crosses in Fig.~\ref{paperpana1:fig:NclNtot}.
The number of \textsl{bound} and \textsl{resonant} states are calculated from the  \textsl{bound} and 
\textsl{resonant} state densities deduced from the microscopic calculation as explained before.
The points in Fig.~\ref{paperpana1:fig:NclNtot} with the lower $N_{cell}$ correspond to isolated nuclei where the 
\textsl{resonant} states are not occupied. 
The dripping point is marked with a vertical line. 
Up to that point one observes a perfect matching between the three quantities represented in Fig.~\ref{paperpana1:fig:NclNtot}
namely $N_{e-cl}$, the number of \textsl{bound} states $N_{bound}$, and the sum of the \textsl{bound}
and \textsl{resonant} states.
When the first neutrons drip out of the clusters 
these three quantities become different.
The number of states in the cluster is slightly larger than the number of bound states, and in general close to the sum of bound and resonant states, 
indicating that 
{resonant} states should be included in the definition of the energy-space cluster.

Notice that the coordinate-space representation induces naturally the idea of an excluded volume for 
the external gas, while the energy-space representation allows the
cluster and the gas to overlap and there is no excluded volume.

To summarize, both representations have their own advantages and disadvantages. They can be exactly mapped onto each other, and we will take advantage of that in the next section. However, the chosen representation has important consequences in terms of allowed excited states and excluded volume, meaning that all these aspects have to be treated consistently  
in the implementation of these representations in a statistical
modeling of supernova matter.

\section{Analytical modeling of the density\label{sec:anmodden}}

In section~\ref{paperpana1:sec:analysis}, the coordinate-space and the energy-space clustering were 
introduced and illustrated in the case of $^{1950}$Sn.
In this section, we make a broader analysis of the results of the fit and we propose an analytical modeling
which reproduces the parameters of the fit with an accuracy of the order of roughly 5\% or better.

In the analytical model the parameters of the Woods-Saxon function, namely the bulk density $\rho_{0}^{WS}$, $R^{WS}$, and
$a^{}$, are expressed as functions of these variables: 
($A_{r-cl}$, $Z_{r-cl}$, $\rho_{gas}$, $\rho_{gas,p}$)
or ($A_{e-cl}$, $Z_{e-cl}$, $\rho_{gas}$, $\rho_{gas,p}$). 
The dependence is not always direct, but goes through the bulk asymmetry $\delta_{r-cl}$ or $\delta_{e-cl}$. 
The populations of the \textsl{e-cluster} and the \textsl{r-cluster} 
are related via Eq.~(\ref{paperpana1:eq:relation}). 
Next we are going to explore these dependencies to complete the modelization of the density profiles. For this purpose we have performed a large number of microscopic calculations for a selection of
nuclei and cells (Ca, Ni, Zr, Sn, Pb), varying the number of neutrons in the cell from stable nuclei
to approximately 3000 with a fixed size of the cell, with radius $R_{cell}$=35~fm.

\subsection{The bulk density}

We first relate the bulk density in the clusters to the saturation property of nuclear matter.
The saturation density $\rho_0$ is defined as the density for which 
the binding energy of nuclear matter is maximized - or where symmetric matter 
is mechanically stable, $P(\rho_0,\delta=0)=0$. 
The saturation density is usually defined in symmetric nuclear matter, for which the isospin asymmetry
$\delta=(N-Z)/A=0$.
The saturation density can however be generalized to asymmetric nuclear matter as the density for 
which asymmetric matter is mechanically stable: $P(\rho_0(\delta),\delta)=0$.
Apart from fluctuations arising from quantum shell structure, ordinary isolated nuclei aim at having a
bulk density close to the saturation property of nuclear matter for an asymmetry that corresponds
to their bulk asymmetry.
In the following, it is shown that the bulk density in dilute nuclear clusters also tends to the same
saturation density.

Considering the Generalized Liquid-Drop Model (GLDM)~\cite{Ducoin2010,Ducoin2011}, 
the binding energy is expressed as a systematic analytical expansion around the saturation density
$\rho_0(\delta=0)$ as,
\begin{equation}
B(x,\delta) = \sum_{n\ge 0}\frac{1}{n!}\left( C_{IS,n}+C_{IV,n} \delta^2\right)x^n,
\label{paperpana1:eq:GLDM}
\end{equation}
where $x=(\rho-\rho_0)/(3\rho_0)$ (here $\rho_0=\rho_0(\delta=0)$ and we will shorten the notation in
unambiguous cases), and the coefficients $C_{IS,n}$ and $C_{IV,n}$ stand for the $n$-derivative
of the binding energy with respect to the density $\rho$ in the isoscalar (IS) and isovector (IV) channels.
We have the following relations: $C_{IS,0}=B(0,0)=B_0$ (binding energy), $C_{IS,1}=0$ (pressure), 
$C_{IS,2}=K_{\infty}$ (incompressibility modulus) for the isoscalar channel, and $C_{IV,0}=J_0$ (symmetry
energy), $C_{IV,1}=L$ (slope of the symmetry energy), $C_{IV,2}=K_{sym}$ (curvature of the symmetry
energy) in the isovector channel. 
The nuclear coefficients $C_{IS,n}$ and $C_{IV,n}$ give a characterization of the nuclear properties around
saturation density.
Much effort has therefore been invested in giving accurate experimental values.
For instance, the binding energy is estimated to be $B_0=-16\pm0.5$~MeV, $K_{\infty}=240\pm40$~MeV,
$J_0=32\pm4$~MeV, $L=65\pm20$~MeV~\cite{RS80}. 
For the SLy4 functional, in particular, the following values are reported in \cite{Reinhard:2005nj}:  
$B_0=-15.972$~MeV, $K_{\infty}=229.97$~MeV, $K_{sym}=-119.74$~MeV, $J_0=32$~MeV, $L=45.94$~MeV.

In the following, we will write for the bulk asymmetry parameter 
\[ 
\delta_{r-cl}=1-2\rho_{0,p}/\rho_0 
\] 
or 
\[ 
\delta_{e-cl}=1-2(\rho_{0,p}-\rho_{gas,p})/(\rho_0 - \rho_{gas}) \] 
 inside the cluster
and 
\[ 
\delta_{r-gas}=\delta_{e-gas} = 1-2\rho_{gas,p}/\rho_{gas} 
\]  
in the gas.
We note that the bulk asymmetry parameter is not equal to the total asymmetry 
\[ I_{r-cl}=1-2Z_{r-cl}/A_{r-cl} \]  or 
\[ I_{e-cl}=1-2Z_{e-cl}/A_{e-cl} \]   
and is more complicated to calculate, 
because of the presence of a neutron or proton skin at the surface.
A model for the bulk asymmetry parameter is presented in section~\ref{paperpana1:section:delta}.

Setting the first density-derivative of Eq.~(\ref{paperpana1:eq:GLDM}) equal to zero, 
the saturation density in asymmetric matter $\rho_0(\delta_{r-cl})$ can be related to the nuclear coefficients
$C_{IS,n}$ and $C_{IV,n}$.
Limiting the series expansion in Eq.~(\ref{paperpana1:eq:GLDM}) to $n=2$, it is found indeed that
\begin{equation}
\rho_0^{GLDM, n=2}(\delta_{r-cl}) = \varrho_0 \left( 1-\frac{3 L \delta_{r-cl}^2}{K_\infty+K_{sym} \delta_{r-cl}^2 }  \right) ,
\label{paperpana1:eq:rho0}
\end{equation}
where $\varrho_0$ the saturation density of symmetric matter. 
A lowest-order expression can be obtained with the limitation to $n=1$ in Eq.~(\ref{paperpana1:eq:GLDM}), 
\begin{equation}
\rho_0^{GLDM, n=1}(\delta_{r-cl}) = \varrho_0 \left( 1-\frac{3 L}{K_\infty} \delta_{r-cl}^2 \right) ,
\label{paperpana1:eq:rho02}
\end{equation}
as it was already obtained in Ref.~\cite{Danielewicz2009}, expression (64).
Notice that, since Eqs.~(\ref{paperpana1:eq:rho0}) and ~(\ref{paperpana1:eq:rho02}) are obtained from
considerations in nuclear matter, the density and asymmetry shall be those of the equivalent nuclear matter system,
which in our notation means $\rho_{r-cl}$ and $\delta_{r-cl}$.

\begin{figure*}[t]
\includegraphics[width=0.7\textwidth]{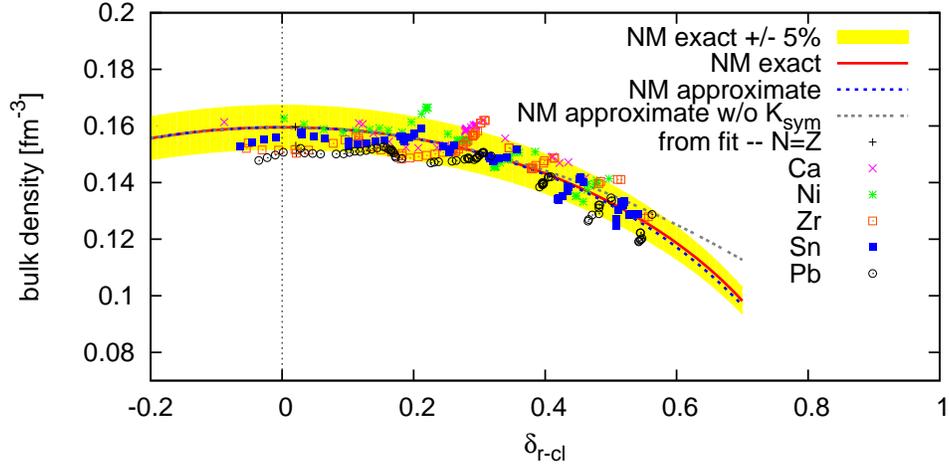} 
\caption{Central (bulk) density of clusters from the fits of microscopic density profiles (points), compared with the saturation density 
of nuclear matter at a given asymetry, given by: an exact numerical calculation, solid red line; 
approximation (\ref{paperpana1:eq:rho0}), short-dashed blue line; and approximation (\ref{paperpana1:eq:rho02}), dotted gray line.}
\label{paperpana1:fig:rho0ofd}
\end{figure*}

The accuracy of the expression~(\ref{paperpana1:eq:rho0}) can be estimated from the comparison to 
microscopic calculations.
The bulk density and asymmetry are extracted from the microscopic calculation as the result of the fit described in
section~\ref{paperpana1:sec:analysis}. 
The so-obtained microscopic bulk density $\rho_0$ is shown in Fig.~\ref{paperpana1:fig:rho0ofd} 
as function of the respective bulk asymmetry 
$\delta_{r-cl}$, where the various types of symbols are associated to various 
isotopic chains.
These are compared with the 
exact result for the SLy4 Skyrme functional and with the 
analytical expression~(\ref{paperpana1:eq:rho0}) $\rho_0^{GLDM, n=2}(\delta_{r-cl})$, 
where the nuclear coefficients $L$, $K_{\infty}$, and $K_{sym}$ are set to be that of SLy4. 
The nuclear coefficients are given, for instance, in Table I of Ref.~\cite{Ducoin2011}.
The yellow band around the exact result expression represents a variation of $\pm$5\%.
The analytical expression $\rho_0^{GLDM, n=2}(\delta)$ reproduces the exact result accurately up to large asymmetries, 
while  
the analytical expression $\rho_0^{GLDM, n=1}(\delta)$ is only reliable at moderate asymmetries.  
The microscopic bulk densities follow the nuclear-matter saturation curve within a deviation of the order of 5\%.
Close to symmetry, the microscopic bulk density is systematically lower than $\varrho_0=\rho_0(\delta =0)$ mostly due to the Coulomb interaction.

\begin{figure*}[t] 
\includegraphics[width=0.97\textwidth]{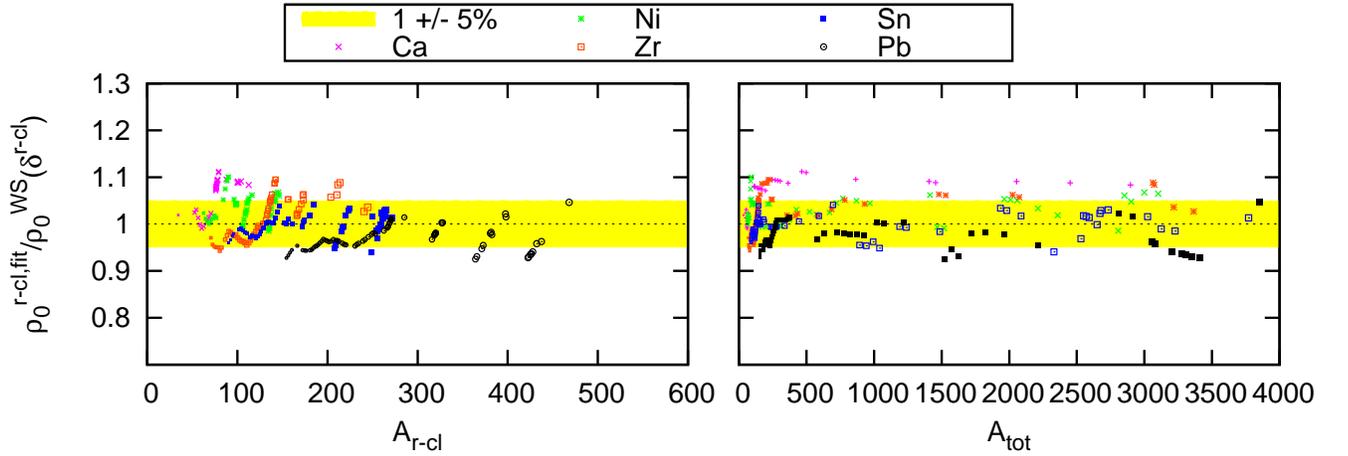} 
\caption{Comparison between the values of the bulk density as obtained from the fits and from the saturation property of nuclear matter, 
as a function of the \textsl{r-cluster} population or the total amount of nucleons in the cell. 
\label{paperpana1:fig:rho0more} } 
\end{figure*} 

In order to analyze whether the transition between isolated nuclei and cells plays a role, the ratio of the microscopic 
bulk density $\rho_0$ over the analytical model $\rho_0^{GLDM, n=2}(\delta)$ are shown
in Fig.~\ref{paperpana1:fig:rho0more} with respect to the number of nucleons inside the coordinate-space
cluster $A_{r-cl}$ (top panel) and with respect to the total number of nucleon in the cell $A_{tot}$
(bottom panel).
It is clear that the analytical model $\rho_0^{GLDM, n=2}(\delta)$ gives an equally accurate
estimation of the microscopic bulk density  $\rho_0$ for both isolated nuclei ($A_{tot} \lesssim 50-300$)
and nuclei embedded in the gas ($A_{tot} \gtrsim 50-300$).

\subsection{The bulk asymmetry parameter}
\label{paperpana1:section:delta}

Knowledge of the neutron and proton density profiles requires the knowledge of the Woods-Saxon parameters
$\rho_{0,q}$,  which are related to the bulk density and the asymmetry parameter $\delta_{r-cl}$ as
\begin{eqnarray}
\rho_{0,n} &=& \frac{1+\delta_{r-cl}}{2}\rho_0(\delta_{r-cl}) \\
\rho_{0,p} &=& \frac{1-\delta_{r-cl}}{2}\rho_0(\delta_{r-cl}).
\end{eqnarray}
A model to estimate the isospin asymmetry $\delta_{r-cl}$ is therefore needed.

In isolated nuclei, the bulk asymmetry $\delta_{e-cl}=\delta_{r-cl}$ and it is slightly different from the global
asymmetry $I_{e-cl}=(A_{e-cl}-2Z_{e-cl})/A_{e-cl}$ due to two effects: first the Coulomb interaction
pushes protons to the surface of heavy nuclei, and the symmetry energy contribute to
produce proton or neutron skins at the surface of nuclei.
The relation between $I_{e-cl}$ and $\delta_{e-cl}$ has been obtained from a liquid-drop 
model~\cite{Myers1980,Centelles1998,Warda2009}
\begin{equation}
\delta_{e-cl} = \frac{I_{e-cl}+\frac{3 a_C}{8 Q} \frac{(Z_{e-cl})^2}{(A_{e-cl})^{5/3}}} {1+\frac{9 J_0}{4 Q} \frac{1}{(A_{e-cl})^{1/3}}},
\label{paperpana1:eq:deltacl}
\end{equation}
where $Q$ is the surface stiffness coefficient extracted from a semi-infinite nuclear matter calculation
and $a_C$ the Coulomb
parameter defined as $a_C=3e^2/(5 r_0)$.
We have $r_0^3=3/(4\pi\rho_0(0))$.
In Ref.~\cite{Warda2009}, a correlation pattern has been obtained for a large number of
nuclear models (Skyrme, Gogny, RMF) between the symmetry energy $J_0$, the slope of the 
symmetry energy $L_0$ and $Q$ as $L_0=144.5 J_0/Q - 55.5$~MeV. 
Using this relation, the value for $Q$ can be estimated for each nuclear interaction.

The bulk asymmetry inside the \textsl{r-cluster}, 
$\delta_{r-cl}=1-2\rho_{0,p}/\rho_0$,  can be decomposed into the asymmetry of the
\textsl{e-cluster} 
$\delta_{e-cl}$ weighted by the fraction, 
$x_{e-cl}=(\rho_0-\rho_{gas})/\rho_0$, of the \textsl{e-cluster} in the \textsl{r-cluster} 
plus the asymmetry in the \textsl{e-gas} $\delta_{e-gas}$ weighted by the 
fraction $x_{e-gas}=\rho_{gas}/\rho_0$ of the \textsl{e-gas} 
inside the \textsl{r-cluster}, namely 
\begin{equation}
\delta_{r-cl} = \left( 1-\frac{\rho_{gas}}{\rho_0(\delta_{r-cl})}\right) \delta_{e-cl}
+\frac{\rho_{gas}}{\rho_0(\delta_{r-cl})} \delta_{gas} 
\label{paperpana1:eq:deltain}
.\end{equation}
The asymmetry of the gas $\delta_{gas}$ is simply related to the variables of the model,
$\delta_{gas}=(\rho_{gas,n}-\rho_{gas,p})/\rho_{gas}$, and is independent of the representation. 
The bulk density is given by expression~(\ref{paperpana1:eq:rho0}), and the
cluster asymmetry $\delta_{e-cl}$ is obtained from Eq.~(\ref{paperpana1:eq:deltacl}).
Notice that $\rho_0$ is a function of $\delta_{r-cl}$ and therefore 
Eq.~(\ref{paperpana1:eq:deltain}) is a self-consistent equation, to be solved by iteration
as we will discuss later on. 

\begin{figure*} [t]
\includegraphics[width=0.9\textwidth]{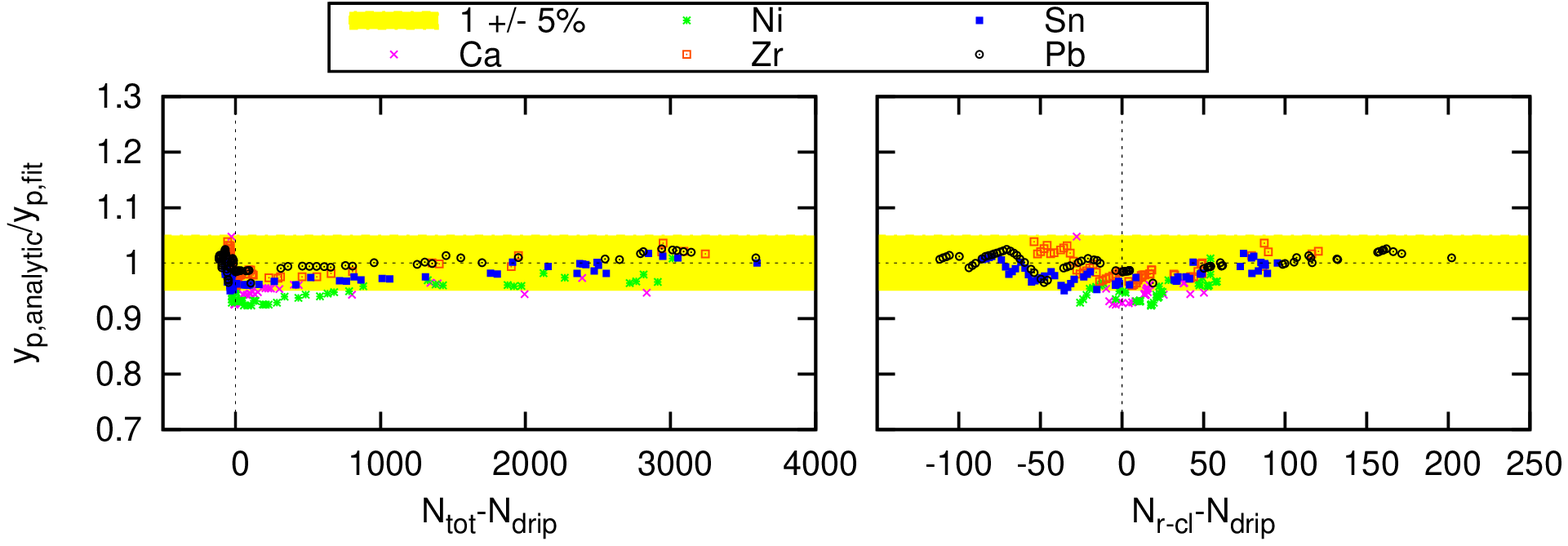} 
\caption{Quality of our approximation for the proton fraction in the interior of the cluster - equivalently, the bulk asymmetry. 
The ratio of the proton fraction, computed via the analytical model, and the value fitted to the microscopic calculations, as a function of: 
the neutron excess or deficiency in the cell with respect to the drip line (left); 
and the neutron excess or deficiency in the \textsl{r-cluster} (right).  
\label{paperpana1:fig:yp0}} 
\end{figure*}

The bulk proton fraction $y_{p,r-cl}$ is obtained from the asymmetry parameter $\delta_{r-cl}$ as 
$y_{p,r-cl}=(1-\delta_{r-cl})/2$.
The ratio of the proton fraction $y_{p,r-cl}$ obtained from the analytical model over the 
microscopic one is shown in Fig.~\ref{paperpana1:fig:yp0} with respect to the total number 
of neutrons $N_{tot}$ shifted by the number of neutrons at the drip-line for each isotopic 
chain $N_{drip}$ (top panel), and with respect to the number of neutrons inside the 
coordinate-space cluster $N_{r-cl}$ shifted by $N_{drip}$.
The results justify our approach. 
A deviation observed arount the drip point is cured as more neutrons are added to the gas. 

One might consider to apply expression 
(\ref{paperpana1:eq:deltacl}) 
to the \textsl{r-cluster} instead. 
In such a case, we found that, compared to the microscopic calculations, 
the asymmetry is overestimated as the density of the neutron gas increases.

\subsection{The surface diffuseness}

\begin{figure*} [t]
\includegraphics[angle=-90,width=0.8\textwidth]{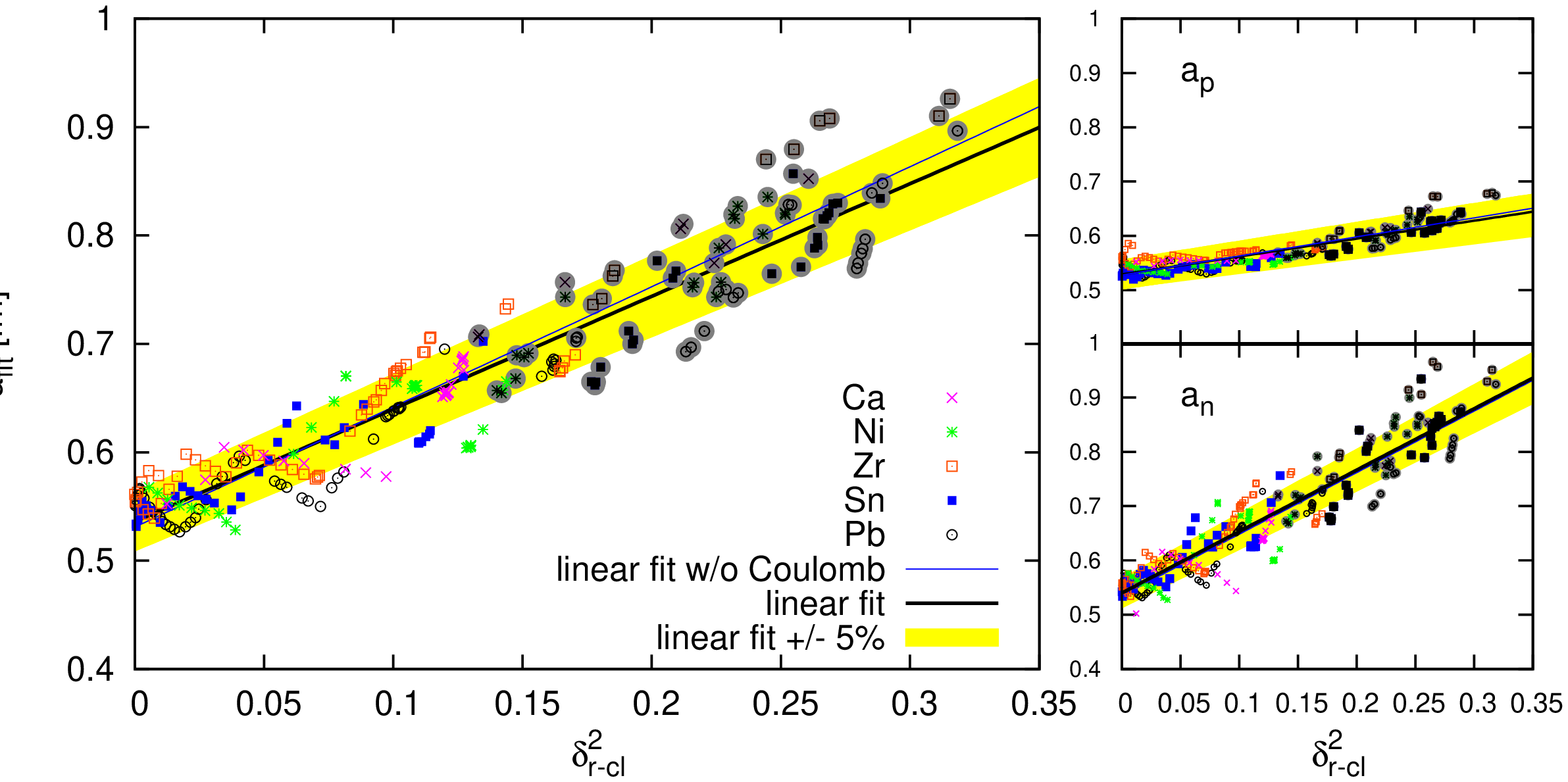} 
\caption{Linear fit of the diffuseness parameter corresponding to: the nucleon density profile (left); 
the proton density profile (right-top); 
and the neutron density profile (right-bottom). 
Colored shapes correspond to isolated clusters, while black shapes in gray disks correspond to clusters in the presence of a gas. 
\label{paperpana1:fig:adiffit}} 
\end{figure*}

In ordinary nuclei, the surface diffuseness $a$ is of the order of $0.55\pm0.05$~fm~\cite{BM75} and this value
is different for light nuclei and heavier stable nuclei.  
There is therefore a dependence on the mass of the nuclei. 
We now turn to show that this mass dependence can be largely absorbed in a dependence on isospin, both in the case of isolated nuclei and nuclei embedded in a gas~\cite{Douchin2000419}. 
The set of nuclear systems studied in this work includes a few neutron-deficient nuclei, stable nuclei and neutron-rich nuclei or in-medium clusters. 
In Fig~\ref{paperpana1:fig:adiffit} we show the surface diffuseness obtained 
from the fit to microscopic density profiles for the same set of nuclear systems as in Fig.~\ref{paperpana1:fig:rho0ofd}. 
The diffuseness of the nucleon density is shown on the left panel. 

An almost linear correlation between the surface diffuseness $a_n$ and the bulk asymmetry $(\delta_{r-cl})^2$
is apparent 
and independent from whether we consider isolated nuclei (colored shapes) or clusters embedded in a gas (black shapes in gray disks). 
It has therefore been fitted as,
\begin{equation}
a = \alpha + \beta \delta_{r-cl}^2 \quad \text{[fm]}.
\label{paperpana1:eq:fitdif}
\end{equation}
The parameters $\alpha$ and $\beta$ 
are given in Table~\ref{paperpana1:tbl:adiffit} considering 
various sets of nuclei in the fit: Ca, Ni, Zr, Sn, Pb, and finally taking into account all the nuclei in the set.
Linear fits were performed for the proton and neutron diffuseness parameters too, 
\begin{equation} 
a_p=\alpha_p + \beta_p \delta_{r-cl}^2 \, ; \, 
a_n=\alpha_n + \beta_n \delta_{r-cl}^2 \, \text{[fm]}  
\label{eq:adiffpn} 
\end{equation} 
and the results are also given in Table~\ref{paperpana1:tbl:adiffit}. 
In all cases the standard deviation 
\[ 
\sigma = \sqrt{\frac{1}{N-1} \sum_{i=1}^N [ a_i - (\alpha + \beta \delta_{r-cl}^2)]^2 } 
, \] 
where $N$ the number of points in each sample and $a_i$ the diffuseness, at each point, from the microscopic fit, 
is equal to about 0.2 for $a_p$ and 0.3 for $a, a_n$. 
If the fit is performed by excluding $N<Z$ nuclei from the sample, very similar results are obtained, 
namely $\alpha = 0.53$, $\beta = 1.06$, $\alpha_p = 0.53$, $\beta_p = 0.35$, $\alpha_n = 0.54$, $\beta_n = 1.14$. 

The various parameters are very stable and depend only weakly on the
chosen set of input.
This indicates that $\alpha$, $\beta$ and the proton and neutron counterparts 
might be related to general properties of the nuclear interaction.
For comparison, we have performed a fit also for a similar set of calculations using the LNS interaction~\cite{Cao2006b}. 
We found in that case $\alpha = 0.53$ and $\beta = 0.98$, i.e., a somewhat smaller diffuseness, especially for the neutron-rich systems. 
However, these values are still compatible with the ones obtained with SLy4, considering the scattering of pseudo-data points, 
quantified by $\sigma$. 
All values are compatible also with the values extracted from experimental data. 


We notice from the values discussed above and from the right panels of Fig.~\ref{paperpana1:fig:adiffit}  
that the diffusness parameters of the proton and neutron density profiles are different,  
in concordance with a recent study of the difference between the proton and neutron surface thickness~\cite{Warda2009}. 
The two possible reasons are the presence 
of a neutron or proton skin at large asymmetries and the Coulomb interaction spoiling the symmetry between the species. 
The proton diffuseness 
$a_p$ is in general lower than the neutron diffuseness $a_n$, except for the smaller bulk asymmetries, which 
correspond to neutron-deficient nuclei. 
The difference increases for neutron-rich systems.  
We have checked that the strongest effect is due to the nuclear interaction. 
We have performed calculations and fits by switching off the Coulomb interaction and obtained very similar results, as shown in Fig.~\ref{paperpana1:fig:adiffit}. 
This means that in proton-rich systems (negative bulk asymmetry) the role of the protons and the neutrons would be roughly reversed, i.e., 
the proton diffuseness parameter would be larger.

\begin{table*} [t]
\setlength{\tabcolsep}{.15in}
\renewcommand{\arraystretch}{1.1}
\begin{tabular}{c|cccccc} 
\hline
  parameters & Ca  & Ni & Zr & Sn & Pb & all sets \\ 
 \hline \hline 
   $\alpha$ &     0.52    &     0.53    &     0.54    &     0.53    &     0.53    &     0.54     \\ 
    $\beta$ &     1.23    &     1.09    &     1.19    &     1.01    &     0.96    &     1.04     \\ 
 \hline 
 $\alpha_p$ &     0.53    &     0.52    &     0.54    &     0.52    &     0.53    &     0.53     \\ 
  $\beta_p$ &     0.36    &     0.37    &     0.34    &     0.35    &     0.32    &     0.33     \\ 
 \hline 
 $\alpha_n$ &     0.51    &     0.54    &     0.54    &     0.54    &     0.54    &     0.54     \\ 
  $\beta_n$ &     1.29    &     1.19    &     1.30    &     1.11    &     1.03    &     1.13      \\
 \hline 
\end{tabular} 
\caption{Parameters in Eq.~(\ref{paperpana1:eq:fitdif}) from linear fits to various sets of nuclear systems:  
Ca, Ni, Zr, Sn, Pb and from all sets.}
\label{paperpana1:tbl:adiffit}
\end{table*} 

Note that for symmetric systems, and ignoring the effect of the Coulomb interaction, the diffuseness parameter should not depend on the nucleon species. 
All our results are consistent with a rounded value $\alpha = \alpha_p = \alpha_n = 0.53$. 
A dedicated investigation of the dependance of all the parameters on the interaction properties would be worth undertaking in the future. 

\subsection{The radius of the cluster}

The radius parameter $R^{WS}$ entering the density profile is defined  
as:
\begin{equation} 
R^{WS} = R^{HS} \left[ 1 - \frac{\pi^2}{3} \left(\frac{a}{R^{HS}}\right)^2 \right]  
\label{paperpana1:eq:radiusws}
\end{equation} 
with the help of the equivalent homogeneous-sphere value, which is given by: 
\begin{equation}
R^{HS} = \left( \frac{3 V^{HS}}{4\pi}\right)^{1/3}.
\label{paperpana1:eq:radius}
\end{equation}
The equivalent homogeneous-sphere volume, $V^{HS}$, is simply expressed as 
\begin{equation}
V^{HS} = \frac{A_{r-cl}}{ \rho_0(\delta_{r-cl})} \equiv V_{cl} 
\label{paperpana1:eq:volume}
\end{equation} 
and defines the volume of the cluster, $V_{cl}$. 

\begin{figure} [t]
\includegraphics[angle=-90,width=0.45\textwidth]{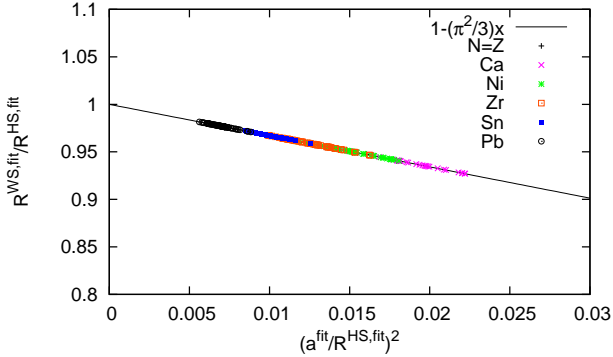} 
\caption{Quality of the relation (\ref{paperpana1:eq:radiusws}).  
\label{paperpana1:fig:radwsmic}} 
\end{figure} 

Eq. (\ref{paperpana1:eq:radiusws})  is  a series expansion in the small parameter $a/R^{HS}$.
In order to check the accuracy of Eq.~(\ref{paperpana1:eq:radiusws}) the ratio $R^{WS}/R^{HS}$  
is shown in Fig.~\ref{paperpana1:fig:radwsmic} as a function of $(a/R^{HS})^2$ and for
the whole set of nuclei considered in this work.
The various symbols correspond to the isotopic chains for Ca, Ni, Zr, Sn, and Pb, while 
the expression~(\ref{paperpana1:eq:radiusws}) is represented by the solid line.
The matching between the symbols and the solid line justify the truncation in
Eq.~(\ref{paperpana1:eq:radiusws}).

\begin{figure*} [t]
\includegraphics[width=0.9\textwidth]{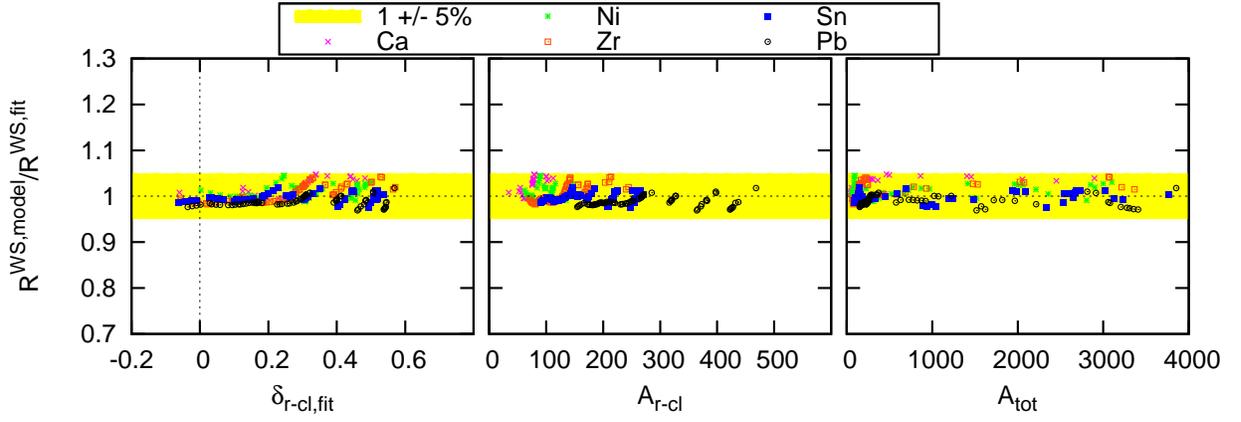} 
\caption{Comparison between the cluster radius determined through the analytical model and the radius obtained from the microscopic density profiles, 
as a function of relevant quantities, namely: 
the bulk asymmetry of the \textsl{r-cluster}; the population of the \textsl{r-cluster}; and the total number of particles in the cell.} 
\label{paperpana1:fig:radws} 
\end{figure*} 

A parameterization of the radius of the Woods-Saxon profile is obtained by injecting 
expressions~(\ref{paperpana1:eq:rho0}), (\ref{paperpana1:eq:radius}), and (\ref{paperpana1:eq:volume}) 
inside Eq.~(\ref{paperpana1:eq:radiusws}).
This model is compared to the Woods-Saxon radius in Fig.~\ref{paperpana1:fig:radws} where
the ratio of the analytical radius~(\ref{paperpana1:eq:radiusws}) over the Woods-Saxon radius
from the microscopic calculations is represented. 
For the completeness of the analysis, the ratio is represented with respect to various
quantities: $\delta_{r-cl}$, $A_{r-cl}$, and $A_{tot}$.
The same set of nuclei as in Fig.~\ref{paperpana1:fig:rho0ofd} is taken into account.
The yellow band give a deviation of $\pm5$\% around the analytical expression.
The analytical radius~(\ref{paperpana1:eq:radius}) gives a good estimation of the Woods-Saxon radius.

\subsection{The analytical model in practice}
\label{paperpana1:sec:themodel}

We can now summarize the basic equations in the model and how they are used in practice. The nucleon density profile is determined by Eqs.~(\ref{paperpana1:eq:ws1}),(\ref{paperpana1:eq:ws2}),(\ref{paperpana1:eq:rhocell}) or, equivalently, 
Eqs.~(\ref{paperpana1:eq:wscluster}),(\ref{paperpana1:eq:wsgas}),(\ref{paperpana1:eq:rhocelle}). 
The variables on which the Woods-Saxon parameters depend are chosen to be the composition of the cluster, which depends on the representation,  and the gas density, which does not. In particular, the variables are either 
($A_{r-cl}$, $Z_{r-cl}$, $\rho_{gas}$, $\rho_{gas,p}$) for the \textsl{r-cluster}, 
or ($A_{e-cl}$, $Z_{e-cl}$, $\rho_{gas}$, $\rho_{gas,p}$) for the \textsl{e-cluster}. 
Equivalently, the gas asymmetry $\delta_{gas}=1-2\rho_{gas,p}/\rho_{gas}$ may replace $\rho_{gas,p}$ as a variable.  
The corresponding density profiles are given by Eqs.~(\ref{paperpana1:eq:ws2}) and (\ref{paperpana1:eq:wscluster}), respectively. 
For the various parameters we now have: 
\begin{equation}
\rho_0(\delta_{r-cl}) = \varrho_0 \left( 1-\frac{3 L}{K_\infty+K_{sym} \delta_{r-cl}^2 } \delta_{r-cl}^2 \right), \label{paperpana1:eq:eq1}
\end{equation} 
\begin{equation} 
a^{} = \alpha + \beta \delta_{r-cl}^2,\label{paperpana1:eq:eq5}
\end{equation} 
\begin{equation} 
R^{WS} = R^{HS} \left[ 1 - \frac{\pi^2}{3} \left(\frac{a}{R^{HS}}\right)^2 \right] 
\end{equation} 
with 
\begin{equation} 
R^{HS} = \left( \frac{3 A_{r-cl}}{4\pi\rho_0(\delta_{r-cl})}\right)^{1/3}.\label{paperpana1:eq:eq7}
\end{equation}
The \textsl{r-cluster} asymmetry entering the above expressions is obtained self-consistently with the help of Eq.~(\ref{paperpana1:eq:eq1}) and the relations 
\begin{eqnarray} 
A_{e-cl} &=& \left(1-\frac{\rho_{gas}}{ \rho_0(\delta_{r-cl})} \right) A_{r-cl};  
\label{paperpana1:eq:eq4}\\
Z_{e-cl} &=& \left(1-\frac{\rho_{gas,p}}{ \rho_{0,p}(\delta_{r-cl})} \right) Z_{r-cl};  
\label{paperpana1:eq:eq4z}\\
\delta_{e-cl} &=& \frac{I_{e-cl}+\frac{3 a_C}{8 Q} \frac{(Z_{e-cl})^2}{(A_{e-cl})^{5/3}}} {1+\frac{9 J_0}{4 Q} \frac{1}{(A_{e-cl})^{1/3}}},\label{paperpana1:eq:eq2}\\
\delta_{r-cl} &=& \left( 1-\frac{\rho_{gas}}{\rho_0(\delta_{r-cl})}\right) \delta_{e-cl} +\frac{\rho_{gas}}{\rho_0(\delta_{r-cl})} \delta_{gas} \label{paperpana1:eq:eq3}
.\end{eqnarray} 
The procedure depends on the chosen variables. If ($A_{r-cl}$, $Z_{r-cl}$, $\rho_{gas}$, $\rho_{gas,p}$) are given, we proceed as follows. We start with the initial estimate $\delta_{r-cl (0)} = 1-2Z_{r-cl}/A_{r-cl}$; we use Eq.~(\ref{paperpana1:eq:eq1}) to estimate $\rho_0^{(0)}$; then Eqs.~(\ref{paperpana1:eq:eq4}), (\ref{paperpana1:eq:eq4z}) estimate the \textsl{e-cluster} population and next Eq.~(\ref{paperpana1:eq:eq2}) its asymmetry; Eq.~(\ref{paperpana1:eq:eq3}) gives a new estimate for the asymmetry, $\delta_{r-cl (1)}$; and so on until convergence. 
If ($N_{e-cl}$, $Z_{e-cl}$, $\rho_{gas}$, $\rho_{gas,p}$) are given, one readily obtains $\delta_{e-cl}$ from Eq.~(\ref{paperpana1:eq:eq2}) and can proceed as follows. An initial estimate for $\rho_0^{WS}$ is the saturation density of symmetric matter, $\rho_0^{(0)} = \varrho_0$. The \textsl{r-cluster} asymmetry is then estimated from Eq.~(\ref{paperpana1:eq:eq3}). This leads to a new estimate $\rho_0^{(1)}$ via Eq.~(\ref{paperpana1:eq:eq1}). 
And so on until convergence. 

The parameters determining the proton and neutron density profiles are readily obtained. 
Proton and neutron bulk densities are determined by the total bulk density and the bulk asymmetry. 
Along with the numbers of neutrons and protons in the cluster they yield the respective radii.  
Diffuseness parameters are given by the approximation (\ref{eq:adiffpn}), with $\alpha_{p,n}$ and $\beta_{p,n}$ 
possibly depending on the interaction, but in general close to the values given in Table~\ref{paperpana1:tbl:adiffit}, 
as already discussed. 
Note that for proton-rich systems ($\delta_{r-cl}<0$) the proton and neutron diffuseness parameters are exchanged with respect 
to the values given in the Table. 
 
In applications relying on an excluded volume, care must be taken to define the volume of the cluster (both protons and neutrons) uniquely. 
A straightforward strategy is to set the proton and neutron \textsl{r-gas} inner radius equal to the total nucleon \textsl{r-gas} inner radius. 
The proton and neutron density profiles would still be determined by the above relations, but a reshuffling of nucleons between the gas and the cluster population might be necessary. 
In our numerical calculations this ambiguity is avoided because there is no proton gas and we do not treat the neutron density profile as independent, see Eqs.~(\ref{paperpana1:eq:rhospacen1}), (\ref{paperpana1:eq:nrcl1}). 

\section{Energy of a cluster in a gas\label{sec:energy}} 

We now turn to the energies of clusters embedded in a nucleon gas. 
We will define these energies based on our two 
representations for dilute clusters
and then 
study them with the help of the analytic density profiles and the local density approximation (LDA).

The LDA allows us to go beyond the restricted domain of Hartree-Fock, namely to study combinations of cluster and gas compositions that would not be generated through energy minimization. 
It is well known from present EoS models that a very large distribution of clusters of any size and isospin is expected in the different thermodynamic conditions relevant for supernova physics. 
Information on what are essentially excited configurations is thus mandatory for the treatment of clusterized matter at finite temperature. 
In fact, we will show that relying solely on Hartree-Fock or, in general, variational calculations (similarly: existing nuclei) can be highly misleading when attempting to generalize the results to more exotic systems. Such results remain of value at certain domains of low temperature, but, in general, one should keep in mind the associated limitations when extrapolating. 

With practical applications in mind, in particular NSE implementations~\cite{Raduta2010}, where the energetics of the gas and of the cluster (whether in energy or coordinate representation) are treated separately, we will partition the total nuclear energy of the Wigner-Seitz cell into the nuclear energy of the cluster and that of the gas. In particular, the energy due to the interaction of the two coexisting systems will be conveniently assigned to the cluster. The gas can then be treated as a uniform system - for example, as is currently done in the statistical model of Refs.~\cite{Raduta2009,Gulminelli2012}, within the temperature-dependent Hartree-Fock method. Then we need only be concerned with the energetics of the cluster. 

Our first task is therefore to disentangle the energy of the cluster from that of the gas. 
For that, let us call $E_{cell}$ the nuclear energy of the whole system $(A_{r-cl},Z_{r-cl},\rho_{gas},\delta_{gas},R_{cell})$, as calculated from a 
microscopic approach. Our aim being to describe arbitrary configurations, that is excited states
as well as ground states, this is given by the LDA approach in terms of the analytical Woods-Saxon density profiles, developed above. 
The total energy can be partitioned, without introducing any ambiguity, as the sum of the bulk energy of the \textsl{r-gas}, the bulk energy of the \textsl{r-cluster}, and a correction $\delta E$, which we will identify as the surface energy: 
\begin{eqnarray} 
E_{cell} &=& 
(V_{cell} - V_{cl}) \varepsilon (\rho_{gas},\delta_{gas}) 
\nonumber \\ 
&& + 
V_{cl} \varepsilon (\rho_0,\delta_{r-cl}) 
+ 
\delta E . 
\end{eqnarray} 
Here $\varepsilon (\rho , \delta )$ is the energy density of homogeneous matter at density $\rho $ and asymmetry $\delta $. 
In the Skyrme model it equals the energy density of  
Eq.~(\ref{eq:skyrmeLDA}), with the gradient terms vanishing.  
In the above, $V_{cl}=\frac{4}{3}\pi R_{HS}^3$ is the volume of the cluster and 
$V_{cell}=\frac{4}{3}\pi R_{cell}^3$ 
is the volume of the cell. 

As already argued, we may absorb $\delta E$ into the energy of the cluster, $E_{r-cl}$, and henceforth be concerned only with this quantity, i.e.,  
\begin{equation} 
E_{r-cl} = E_{cell} - 
(V_{cell} - V_{cl}) \varepsilon (\rho_{gas},\delta_{gas})  
\label{eq:Er-cl-general} 
. 
\end{equation} 
For Skyrme functionals within the LDA, the above yields readily an integro-analytical expression for the energy of a cluster $(A,Z)$ embedded in a gas of given density and asymmetry: 
\begin{eqnarray} 
\lefteqn{E_{r-cl}(A_{r-cl},Z_{r-cl},\rho_{gas},\delta_{gas})  =} 
\nonumber    \\ 
&& 4\pi \int_0^{R_{cell}} \!\!\! \varepsilon ( \rho^{WS}_{cell} (r),\delta_{cell} (r) ) 
r^2 \mathrm{d}r 
\nonumber 
\\ 
&&  - \frc{4}{3}\pi \{ R_{cell}^3 
\nonumber \\ 
&& - [R^{HS}(A_{r-cl},Z_{r-cl},\rho_{gas} ,\delta_{gas})]^3 \} \varepsilon (\rho_{gas} ,\delta_{gas}) 
, 
\label{eq:Er-cl} 
\end{eqnarray} 
where the local density $\rho^{WS}_{cell}$ 
as well as the local asymmetry 
\begin{equation} 
\delta_{cell}(r)=1-2\rho^{WS}_{cell,p}(r)/\rho^{WS}_{cell}(r) 
\end{equation} 
(see also Eqs.~(\ref{paperpana1:eq:rhocell}),(\ref{paperpana1:eq:rhospacep})) 
are determined by $(A_{r-cl},Z_{r-cl},\rho_{gas},\delta_{gas})$ within our analytical model.  
Quite similarly, we readily have an expression for the energy of the \textsl{e-cluster}: 
\begin{eqnarray} 
\lefteqn{E_{e-cl}(A_{e-cl},Z_{e-cl},\rho_{gas},\delta_{gas})  =} 
\nonumber  \\ 
&& 4\pi \int_0^{R_{cell}} \!\! \rho^{WS}_{cell} (r;A_{e-cl},Z_{e-cl},\rho_{gas} ,\delta_{gas}) 
r^2 \mathrm{d}r 
\nonumber  \\ 
&&  - \frc{4}{3}\pi R_{cell}^3 \varepsilon (\rho_{gas} ,\delta_{gas}) 
, 
\label{eq:Ee-cl} 
\end{eqnarray} 
where the cluster-gas interaction energy in the region where they coexist, 
namely $V_{cl}$,  has been absorbed into the energetics of the cluster. 
There is no ambiguity in this choice, since the total energy in the cell is conserved by construction. 
In the above expressions, the radius of the cell $R_{cell}$, chosen randomly and much larger than the cluster's radius, is relevant only to the gas energy and therefore is an auxiliary quantity, not affecting the energetics of the cluster, in either representation. 

Next we will discuss our numerical results based on the above results. 
First, we will discuss the total cluster energy in order to validate the LDA. 
Next, 
we will discuss separately the bulk and surface energy of the \textsl{r-cluster}. 
 
\subsection{Total cluster energy}

\begin{figure} [t]
\includegraphics[width=0.49\textwidth]{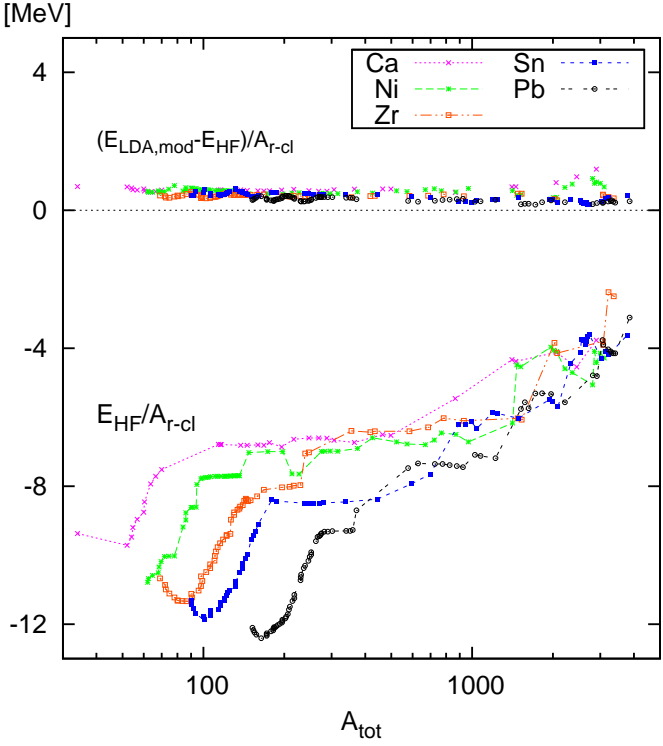} 
\caption{Points connected with lines: Energy of \textsl{r-clusters}, determined by subtracting the \textsl{r-gas} contribution from the total Hartree-Fock energy in the cell. 
The points in the upper part indicate the deviations when the total energy of the cell is computed within the LDA and using the model density profiles. 
\label{fig:justifyLDA} 
}
\end{figure}

In Fig.~\ref{fig:justifyLDA} we show the nuclear part of the cluster energies 
resulting from the Hartree-Fock calculation, as points connected with lines. 
They are computed using Eq.~(\ref{eq:Er-cl-general}) with $E_{cell}$ the Hartree-Fock energy of the cell. 
We notice that the binding energies of the clusters decrease as the number of nucleons in the system increases. 
Indeed, for a given isotopic chain, $A_{tot}$ controls the total number of neutrons. 
The larger its value, the more asymmetric is the cluster that is favored by energy minimization, which is reflected in the decreased binding. 
We will return to the role of the asymmetry when we discuss separately the volume and surface energy. 

We now ask whether the LDA is an acceptable approximation to the microscopic energy. 
Therefore, we will compare the Hartree-Fock energies of the clusters with their LDA energies, computed from Eq.~(\ref{eq:Er-cl-general}) with $E_{cell}$ being replaced by the LDA energy of the cell, using the analytical density profiles - equivalently, Eq.~(\ref{eq:Er-cl}).
The difference of the two energies per particles is also plotted in Fig.~\ref{fig:justifyLDA}. 
It corresponds to the points just above zero in the figure. 
We notice that the difference remains roughly the same for all isotopic chains and $A_{tot}$ (or asymmetries), 
and regardless of the presence or not of a gas,  
at the level of half an MeV per cluster particle. 
This is a very useful outcome: if we are to quantify additive in-medium modifications to nuclear energies, 
the LDA will be as reliable as Hartree-Fock at the very least.  

\subsection{Bulk energy} 

Even within the analytical density profiles in the LDA approximation, the in-medium modified cluster energy functional is a complex integral function of four variables, namely the cluster atomic and mass number Z and A, as well as the density $\rho_{gas}$ and asymmetry $\delta_{gas}$ of the surrounding gas. To simplify this expression and get an insight into its physical meaning, we   
will distinguish between the volume energy and the surface energy of the cluster. 
A separation into surface and volume terms will also help develop formal extentions of the liquid drop model, in the form of, e.g.,  $\delta a_i (\rho_{gas},\delta_{gas})$, where $a_i$ a liquid-drop parameter and $\delta a_i$ its correction, depending on the composition of the gas at the very least. Such additive corrections could be plugged into any mass table or model in an intuitively simple manner.

\begin{figure} [t]
\includegraphics[width=0.48\textwidth]{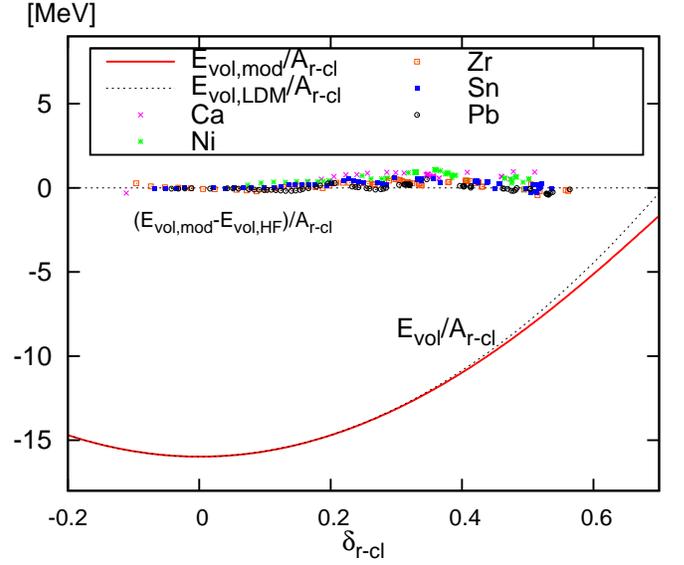} 
\caption{The lines show the volume energy per particle in the \textsl{r-cluster}, 
for the SLy4 functional, 
according to our model (red solid line) 
and according to a liquid-drop model, quadratic in $\delta_{r-cl}$ (dotted line). 
The points show, for the various isotopic chains studied in this work, 
the deviation of our model volume energy per particle 
from the values obtained directly from the 
Hartree-Fock calculations. 
\label{fig:BulkEnergy} 
}
\end{figure}

The bulk (or volume) energy per particle of the \textsl{r-cluster} is defined as 
\begin{equation}
E_{vol}(A_{r-cl},\delta_{r-cl},\rho_{gas},\delta_{gas})=\frac{A_{r-cl}}{\rho_0(\delta_{r-cl})} \varepsilon (\rho_0(\delta_{r-cl}),\delta_{r-cl}) . \label{evol}
\end{equation}
The bulk asymmetry of the cluster $\delta_{r-cl}$, corresponding to given variables $(A_{r-cl},Z_{r-cl},\rho_{gas},\delta_{gas})$, 
can be computed through our analytical model and determines uniquely the bulk density as well. 
The calculation of the bulk energy with the help of the Skyrme energy density is then straightforward. 
The result is a function of the bulk asymmetry, and is shown as the 
solid  line in Fig.~\ref{fig:BulkEnergy}. 
A quadratic function, corresponding to a simple liquid-drop model for the volume energy 
and with parameters which are consistent with the SLy4 functional~\cite{Reinhard2006}, is also plotted (dotted line). 
At low asymmetries the two curves coincide. 
The role of higher-order terms in the asymmetry, 
including those due to the kinetic energy, 
becomes apparent at large asymmetries.  

We take the opportunity to perfom one more test of our analytical modeling of bulk cluster densities. 
In Fig.~\ref{fig:BulkEnergy} the points correspond to the cluster configurations which we obtained within Hartree-Fock. 
They show the deviation between 
a) the bulk energy per particle, calculated through the analytical density model as described above, 
and 
b) 
the microscopic bulk energy per particle, namely the one calculated using those values for the bulk density and asymmetry, which resulted from a fit to the microscopic density profiles. 
We notice that the deviations are consistent with zero and in general there are no systematic deviations as we go to large asymmetries. 
Small deviations corresponding roughly to the drip point are cured at higher asymmetries (cf Sec.~\ref{paperpana1:section:delta}
and Fig.~\ref{paperpana1:fig:yp0}).

\subsection{Surface energy} 

We now proceed to a study of the quantity $\delta E$, 
which is the difference of the total energy and the bulk energy 
and which we will identify as the surface energy. 
We will study this quantity in terms of the \textsl{r-cluster}. 
Switching to \textsl{e-cluster} variables is straightforward, thanks to the geometric relations between the two types of clusters. 

In principle, $\delta E$ depends in unknown ways on the variables $(A_{r-cl},Z_{r-cl},\rho_{gas},\delta_{gas})$. 
If it is a proper surface quantity, however, it should scale as $A_{r-cl}^{2/3}$. 
We furthermore expect that $\delta_{r-cl}$ can replace $Z_{r-cl}$ as the relevant cluster parameter. 
We are then left with the variables $(\delta_{r-cl},\rho_{gas},\delta_{gas})$. 
Regarding the quantity 
\begin{equation} 
s(\delta_{r-cl},\rho_{gas},\delta_{gas}) = 
\frac{\delta E}{A_{r-cl}^{2/3}} 
,
\end{equation} 
the following limiting case should hold: 
\begin{equation} 
s ( \delta_{r-cl},\rho_0(\delta_{r-cl}),\delta_{r-cl}) = 0 
, 
\label{eq:SurfEnLim}
\end{equation}
because in the case where the cluster and the gas have the same density and asymmetry we are in the limit of homogeneous matter and therefore the surface energy should vanish. 
Furthermore, for relatively small absolute asymmetries, 
\begin{equation} 
s ( \delta_{r-cl},\rho_{gas},\delta_{gas}) \approx 
s ( -\delta_{r-cl},\rho_{gas},-\delta_{gas}) 
. 
\label{eq:SurfEnSym}
\end{equation}
The weak violation
of this relation is due to Coulomb effects. 
 
We will now focus on a neutron gas, $\delta_{gas}=1$, though very similar studies can be performed in other cases. 
In short, we are interested in the quantity 
\begin{equation} 
s(\delta_{r-cl},\rho_{gas}) = 
\frac{\delta E}{A_{r-cl}^{2/3}} 
,
\end{equation} 
assuming a neutron gas.  

\begin{figure} [t]
\includegraphics[angle=-90,width=0.45\textwidth]{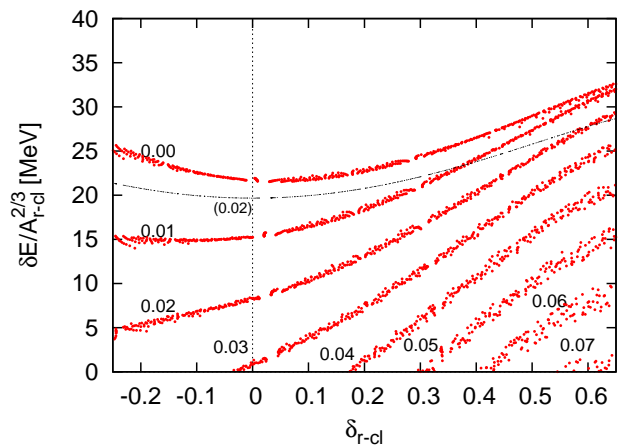} 
\caption{Model surface energy for the SLy4 functional, obtained as described in the text, for given neutron-gas density (values indicated in fm$^{-3}$) as a function of bulk \textsl{r-cluster} asymmetry. 
The secondary line for $\rho_{gas}=0.02$~fm$^{-3}$, marked (0.02), corresponds to the ansatz (\ref{eq:ansatzJM}). 
\label{fig:SurfEnergy_givenrho_paper_one} 
}
\end{figure}

\begin{figure} [t]
\includegraphics[angle=-90,width=0.45\textwidth]{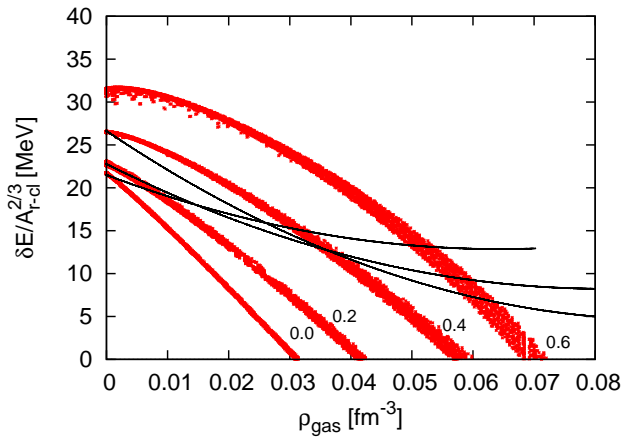} 
\caption{Thick labelled bands of points: Model surface energy for the SLy4 functional, obtained as described in the text, for given bulk \textsl{r-cluster} asymmetry (values indicated) as a function of neutron-gas density. 
Thinner lines: Ansatz (\ref{eq:ansatzPP}), for bulk asymetries equal to 0, 0.2 and 0.4. 
\label{fig:SurfEnergy_givendelta_paper_one} 
}
\end{figure}

In principle, the functional dependence of $s$ on the two variables is not unique. 
Different clusters $(A_{r-cl},Z_{r-cl})$ may correspond to the same combination of $(\delta_{r-cl} , \rho_{gas})$. 
The uniqueness will have to be demonstrated numerically. 
To this end we proceed as follows: 
For given $\rho_{gas}$ we obtain the analytical density profiles corresponding to nuclei with many different combinations of $(A_{r-cl},Z_{r-cl})$ and with a large variation in particle number, namely from 50 to 400 particles. 
This provides us with a value for the bulk asymmetry. 
We also calculate $\delta E $ as already demonstrated. 
We thus obtain, for given $\delta_{r-cl}$ and $\rho_{gas}$, a number of surface-energy results corresponding to clusters of different populations. 
These ``data points" are displayed in Figs.~\ref{fig:SurfEnergy_givenrho_paper_one} and \ref{fig:SurfEnergy_givendelta_paper_one} and discussed below.  

Let us begin with Fig.~\ref{fig:SurfEnergy_givenrho_paper_one}, where 
$s(\delta_{r-cl},\rho_{gas})$ 
is plotted as a function of $\delta_{r-cl}$ for the indicated values of gas density (given in fm$^{-3}$). 
For low gas densities the points form well defined lines, and even at higher gas densities the dispersion of points is moderate. We conclude that the scaling with respect to $A_{r-cl}^{2/3}$ indeed holds to a satisfactory degree. 
This important result means that the definition we have introduced, eq.(\ref{evol}), correctly represents the bulk part of the energy of the cluster. We notice that eq.(\ref{evol}) has no dependence on the external gas.  This means that the presence of a medium does not influence the bulk term, which keeps being defined by the local nuclear matter saturation condition whatever the external medium. This demonstrates that the energy increase of the cluster with increasing neutron gas observed in  Fig.~\ref{fig:justifyLDA} is not an in-medium effect, but simply reflects the decreasing saturation density
(and increasing surface energy, see below) associated to an increasing bulk asymmetry. 

The points forming the line for $\rho_{gas}=0.0$ correspond to isolated nuclei. 
For small asymmetries, a roughly quadratic law is observed. 
Interestingly, the line is convex around zero asymmetry, implying, in a liquid-drop picture, a positive surface-symmetry coefficient, 
in qualitative agreement with certain calculations in semi-infinite matter~\cite{Meyer2003}, but in contrast with many evaluations from the literature~\cite{RPL1983,Danielewicz2003}, including ones based on the SLy4 interaction~\cite{Reinhard:2005nj}. 
Our result, which we found quite robust with respect to approximations involved (for example, the use of LDA instead of Hartree-Fock energies), warrants further investigations in the future. 
To some extent the discrepancies may be due to employing different definitions for the nuclear surface and different partitions between the volume and surface energies~\cite{RPL1983,Centelles1998,Danielewicz2003},  
as well as due to our use of the bulk asymmetry $\delta$ rather than the global asymmetry $I$ as the relevant parameter. 

We observe that, as the neutron-gas density increases, the surface energy decreases. 
On the neutron-rich side this is easy to understand if we consider the limiting case of a gas with the same density and asymmetry as the cluster, defined by Eq.~(\ref{eq:SurfEnLim}). 
Furthermore, 
we observe that the various lines are not symmetric around zero. 
This should be expected, since a proton-rich cluster in a neutron-rich gas is physically a very different entity from a neutron-rich cluster in a neutron-rich gas. 
It is worth noting that the functional dependence at low or negative asymmetries could not have been inferred from our Hartree-Fock calculations, which naturally favor neutron-rich clusters. 
Only the
 analytical model for the density profiles 
 could probe this extended domain. 
The only assumption related to energy minimization is that the bulk density of the cluster equals the saturation density at the given value of asymmetry. 
A different bulk energy would imply compression (or decompression), which would cost too much energy to be 
counter-balanced.
 
Of course, symmetric (not to mention, proton-rich) clusters would not survive in a neutron gas in most situations of interest, in particular low-temperature asymmetric matter in the crust of a neutron star. 
At sufficiently high temperatures, though, such configurations could become active 
and their correct treatment might be important.  

We now turn to Fig.~\ref{fig:SurfEnergy_givendelta_paper_one}, 
where the quantity $s(\delta_{r-cl},\rho_{gas})=\delta E/A_{r-cl}^{2/3}$ is plotted for the indicated values of asymmetry as a function of gas density. 
We notice again that it scales well with $A_{r-cl}^{2/3}$ and it generally decreases for higher gas densities.   
We stress once more that the functional forms for the lower asymmetries would not have been possible to obtain based on an unconstrained variational calculation. 

The functional form of the surface energy obtained here 
goes beyond the somewhat heuristic approaches to the surface energy introduced in Refs.~\cite{Shen1998,Furusawa:2011wh}.  
As a general statement, it appears that simple analytic estimations of surface corrections can be largely inaccurate
and the complete result from the LDA has to be considered.
As an example,
one may suppose that the in-medium corrections can be simply taken into account through the transformation 
from  \textsl{r-clusters} to  \textsl{e-clusters}. Indeed, we have seen that \textsl{e-clusters} can be interpreted as the ensemble of single-particle wave functions (bound and resonant) in the cluster region, with the exclusion of the continuum states that represent the external gas. In this picture one should have: 
\begin{equation}
\delta E =a_s(\delta_{e-cl}) A_{e-cl}^{2/3} , \label{ansatz}
\end{equation}
where $a_s$ can depend on the cluster asymmetry but should not depend on the gas.
Since $A_{e-cl}=(1-\rho_{gas}/\rho_0(\delta_{r-cl}))A_{r-cl}$, 
the uniform-matter limit (\ref{eq:SurfEnLim}) is automatically satisfied.  
Replacing into eq.~(\ref{ansatz}) we get
\begin{equation} 
s(\delta_{r-cl},\rho_{gas}) = \left[ 1-\frac{\rho_{gas}}{\rho_0(\delta_{r-cl})} \right]^{2/3} 
s(\delta_{r-cl},0 ) 
. 
\label{eq:ansatzJM} 
\end{equation} 
The results for $s(\delta_{r-cl}, \rho_{gas}=0.02\textrm{fm}^{-3})$ are shown in Fig.~\ref{fig:SurfEnergy_givenrho_paper_one} (line marked ``(0.02)"). 
For extremely neutron-rich clusters the ansatz works quite well, but, being symmetric around zero asymmetry, it fails to reproduce the correct trends at lower asymmetries. 

Finally, it is worth commenting on an ansatz which we tried at earlier stages of this work~\cite{GMRPO2013}. 
Inspired by the liquid-drop formula and the conditions (\ref{eq:SurfEnLim}),~(\ref{eq:SurfEnSym}), it was proposed:  
\begin{eqnarray} 
s(\delta_{r-cl},\rho_{gas}) 
&=& 
\left(1 - \frac{\rho_{gas}}{\rho_0(\delta_{r-cl})}\right)^2 
\left[ 
a_s + a_{ssym} \delta_{e-cl}^2 
\right] 
\nonumber \\ 
&=& 
a_s\left( 1 - \frac{\rho_{gas}}{\rho_0(\delta_{r-cl})}\right)^2 
\nonumber \\ 
 && + a_{ssym} 
\left( \delta_{r-cl} - \frac{\rho_{gas}}{\rho_0(\delta_{r-cl})}\right)^2 
,
\label{eq:ansatzPP}
\end{eqnarray} 
where $a_s$ and $a_{ssym}$ liquid-drop constants such that the surface energies for $\rho_{gas}=0$ are reproduced. 
The surface-symmetry parameter $a_{ssym}$ was therefore positive. 
An excellent agreement of this ansatz with the Hartree-Fock results of neutron-rich clusters in a neutron gas was observed~\cite{GMRPO2013}. 
We now notice from its analytical form, that this ansatz predicts a larger surface energy for a proton-rich cluster than for a neutron-rich one (in a neutron gas), i.e., the opposite effect than what our LDA approach gave. 
This is not to say that proton-rich clusters in a neutron gas are a particularly relevant configuration, however this outcome demonstrates the risks of relying on Hartree-Fock results to derive general trends.  
The disagreement between our ansatz and the LDA results is further demonstrated in 
Fig.~\ref{fig:SurfEnergy_givendelta_paper_one}. 
The surface energy given by Eq.~(\ref{eq:ansatzPP}) with $a_s=21.5$ and $a_{ssym}=32$ corresponds to the thinner curves (the scaling with $A^{2/3}$ is respected by construction) in Fig.~\ref{fig:SurfEnergy_givendelta_paper_one}. 
The choice of $a_s$ and $a_{ssym}$ is such, that for isolated nuclei, $\rho_{gas}=0$, the LDA and the ansatz give the same result. 
For low gas densities, below $0.02$fm$^{-3}$, there is reasonable agreement.   
It is obvious, however, that there are large deviations in various domains of gas density and asymmetry.

\section{Conclusions\label{sec:concl}}

This work presents a quasi-analytic modelling of the in-medium modifications to cluster density profiles
and energies, when these latter are immersed in a dilute nucleon gas, as it is the case for the stellar matter
which is produced in the core of supernova and in the crust of neutron stars. 

Our approach is valid in principle for all but the lightest nuclei, where there are not enough nucleons to yield saturated matter.  The density profile in a wide number of nuclei and Wigner-Seitz cells, 
as obtained from Hartree-Fock calculations with the SLy4 effective interaction,  
have been analyzed and fitted
through Woods-Saxon functions showing a continuous description from isolated nuclei to nuclei in a medium.

We have discussed two alternative possible representations of clusters inside a medium, depending on the local density
(coordinate-space clusters) or on the energy of the single-particle states (energy-space clusters).
Using these two representation and the mapping between them, we have been able to propose 
an analytical self-consistent model to relate the parameters of the Woods-Saxon 
functions to the global variables of the EoS.

The energies of clusters in a neutron-rich gas were studied with the help of the local-density approximation, which allowed us to investigate configurations which are not explored by the Hartree-Fock (in general, variational) calculations, but are of importance in a finite-temperature stellar environment.

The quality of the LDA approximation was tested on the Hartree-Fock sample. Not surprisingly, this approximation 
systematically overestimated the microscopic result. However, the in-medium cluster energy shift is remarkably well reproduced, showing that the influence of the medium is essentially due to the modification of the local density profile.
We have discussed in detail the in-medium effects due to a pure neutron gas, which is relevant for the physics of the neutron star crust. For this case, we have demonstrated through the comparison with the microscopic calculations that
neither the isoscalar nor the isovector part of the bulk energy of the clusters  in the coordinate space representation are affected by the external gas, and the global bulk energy only depends on the bulk asymmetry of the cluster. As a consequence, the excluded-volume approach appears as a reasonable zero order approximation to account for bulk  in-medium effects. 
An important modification of the surface energy of the clusters is however observed. A decreasing surface tension with increasing density of the gas is observed for all cluster asymmetries.  This leads to a sizeable modification of the cluster energy functional in a dense medium, which will have to be accounted for to have reliable EoS's for astrophysical applications in a near future.

%

In the future, we shall add the Coulomb corrections to the proton mean-field as well as the contribution of the
pairing correlations to the density profiles and the Wigner-Seitz binding energies.

\vspace{2mm} 
{\bf Acknowledgements:} Work supported by the ANR project ``SN2NS: supernova explosions, from stellar core-collapse to neutron stars and black holes".  

\appendix 

\section{Woods-Saxon fit of the microscopic density} 
\label{paperpana1:sec:denfit} 

Here we describe how we fit a microscopic density profile, corresponding to nucleons, protons, or neutrons, 
with a Woods-Saxon analytical 
profile, 
\begin{eqnarray} 
\rho_{cell}^{WS} (r) 
&=& \frac{\rho_0-\rho_{gas}}{1+\exp{[(r-R^{WS})/a]}} + \rho_{gas} 
\\ &=& 
\frac{\rho_0}{1+\exp{[(r-R^{WS})/a]}} + \frac{\rho_{gas}}{1+\exp{[-(r-R^{WS})/a]}}  
.
\end{eqnarray} 
The microscopic density  $\rho_{cell}^{mic}(r_i)$ is given on a radial mesh of N points 
with $r_i=iR_{step}$ spanning a Wigner-Seitz cell of radius $R_{cell}=NR_{step}$. 
During the fit we wish to conserve the total number of particles, 
\begin{equation} 
A^{mic} = 4\pi \left[ \sum_{i=1}^{N-1} r_i^2 \rho_{cell}^{mic} (r_i) R_{step} + \frac{1}{2}r_N^2 \rho_{cell}^{mic} (r_N) \right] R_{step} 
\end{equation} 
within a given tolerence $\delta A $ of, typically, half a particle, 
\begin{equation} 
\begin{array}{c} 
|A^{WS} -A^{mic} | = 
\\  
= \left| 4\pi \left[ \sum_{i=1}^{N-1} r_i^2 \rho_{cell}^{WS} (r_i) + \frac{1}{2}r_N^2 \rho_{cell}^{WS} (r_N) \right] R_{step} -A^{mic} \right| 
\\ 
\leq \delta A . 
\end{array} 
\label{eq:tolerance} 
\end{equation} 
The optimal profile, within this restriction, shall be determined by minimizing the standard deviation 
\begin{equation}
\sigma_A = \sqrt{ \frac{1}{N-1}\sum_{i=1}^N \left( \rho_{cell}^{mic}(r_i)-\rho_{cell}^{WS}(r_i) \right)^2} 
. 
\end{equation}
The parameters to be determined are the bulk and gas densities $\rho_0$ and $\rho_{gas}$, the radius $R^{WS}$ and the diffuseness parameter $a$. 
At the beginning of the fitting procedure the standard deviation 
is initialized at a large number. 

The radius corresponds to the inflection point of the Woods-Saxon profile, 
\begin{equation} 
\left. 
\frc{\mathrm{d}^2}{\mathrm{d}r^2}\rho_{cell}^{WS}(r) 
\right|_{r=R^{WS}} = 0 
\end{equation} 
and can be readily obtained by numerical differentiation of the microscopic density. 
All radial derivatives are calculated at 5-point precision. 
The radius will lie between two points for which the second derivative changes sign. 
Its precise value is obtained by interpolating between the two points. 
We then determine the value of the density as well as its derivative at this point, by interpolation, 
$\rho_{cell}^{mic}(R^{WS})$ and $\mathrm{d}{\rho_{cell}^{mic}}(r)/\mathrm{d}r|_{r=R^{WS}}$, respectively.  
The value of the radius as well as the density and its derivatives are thus determined directly 
from the microscopic density profile and remain fixed during the fit. 

Next we shall vary $\rho_0$ and $\rho_{gas}$ within limiting values, 
$[\rho_{0,\min},\rho_{0,\max}]$ 
and $[\rho_{gas,\min},\rho_{gas,\max}]$ respectively,
determined from the microscopic profile. 
The microscopic density outside the cluster is not perfectly homogeneous, 
but shows weak oscillations and more so towards the radius of the Wigner-Seitz cell. 
The oscillations are largely an artefact of the microscopic calculation and the imposed boundary conditions. 
We therefore determine the limiting acceptable values for $\rho_{gas}$ as the 
minimum and maximum numerical value of $\rho^{mic}(r)$ within a radius $[2R^{WS},R_{cell}]$, 
$\rho_{gas,\min}$ and $\rho_{gas,\max}$, respectively. 

A typical density profile, determined through Hartree-Fock, will show ripples in the interior of a nucleus or cluster. 
Initially we determine the minimum and maximum numerical value of $\rho_{cell}^{mic}(r)$ within a radius equal to $0.7R^{WS}$, 
$\rho_{0,\min ,0.7R^{WS}}$ and $\rho_{0,\max ,0.7R^{WS}}$, respectively. 
If we determine the minimal value of the density up to a value too close to $R^{WS}$, we run the risk of accidentally accepting a too-low value of $\rho_0$. 
On the other hand, we wish to avoid that 
the minimal and maximal values are too-restricted, due to their limitation up to $0.7R^{WS}$. 
Indeed, it is likely that the shoulder of the density profile is critical for a correct determination of $\rho_0$. 
Therefore we extend the allowed interval by considering that, 
for genuine Woods-Saxon profiles, we have: 
\begin{equation} 
\rho_{cell}^{WS} (R^{WS}) = \frac{1}{2} (\rho_0 + \rho_{gas}) \Rightarrow \rho_0 = 2\rho_{cell}^{WS}(R^{WS}) - \rho_{gas} .  
\end{equation} 
Taking this into account, the minumum and maximum acceptable values for $\rho_0$ are determined by 
\begin{equation} 
\begin{array}{c} 
\rho_{0,\min} = \min\{\rho_{0,\min ,0.7R^{WS}},2\rho^{mic}{(R^{WS})}-\rho_{gas, \max} \} 
\\ 
\rho_{0,\max} = \max\{\rho_{0,\max ,0.7R^{WS}},2\rho^{mic}{(R^{WS})}-\rho_{gas, \min} \} 
\end{array} 
\end{equation} 
Finally we vary, within the acceptable intervals and by sufficiently small steps, both $\rho_0$ and $\rho_{gas}$ in nested loops. 
At each step we proceed as follows: the diffuseness parameter is estimated from 
Eq.~(\ref{eq:adiffWS}).  
We can then calculate the number of particles corresponding to the given Woods-Saxon profile. 
If it deviates from the desired value by more than $\delta A$, we reject the current values of $\rho_0, \rho_{gas}, a$ and continue. 
If it is close to the desired value within $\delta A$, we calculate also the 
standard deviation, $\sigma_{loop}$. If this is lower than the stored value of $\sigma$, then $\sigma$ is assigned the new value $\sigma_{loop}$ 
and $\rho_0,\rho_{gas},a$ are assigned the corresponding new values. And so on, until all values have been tried. 
It can happen that no combination of values satisfies the criterion (\ref{eq:tolerance}). 
The reason may be, e.g., an anomalous decrease or increase of the gas density close to the cell radius, due to the boundary conditions. 
Then the Hartree-Fock calculation is rejected. 
This explains the seemingly random gaps in our microscopic results along isotopic chains. 


%

\end{document}